\documentclass[draftcls,onecolumn,twoside,12pt]{IEEEtran}
\usepackage{times}
\usepackage[english]{babel}
\usepackage[latin1]{inputenc}
\usepackage{graphicx}
\usepackage[scanall]{psfrag}
\usepackage{latexsym}
\usepackage{amsfonts}
\usepackage{amssymb}
\usepackage[mathcal]{euscript}
\usepackage{amsmath}
\usepackage{colordvi}
\usepackage{verbatim}
\usepackage{threeparttable}
\usepackage{cite}
\usepackage[table]{xcolor}
\usepackage{algorithm}
\usepackage{algpseudocode}
\usepackage{subfig}
\usepackage{mathtools}

\newcommand\cooloverone[2]{\mathrlap{\smash{\overset{\raisebox{0.8em}{\mbox{$#1$}}}{\phantom{%
    \begin{matrix} #2 \end{matrix}}}}}#2}

\newcommand\coolleftbraceone[2]{%
#1\vphantom{\begin{matrix} #2 \end{matrix}}}

\newtheorem{definition}{Definition}
\newtheorem{proposition}{Proposition}
\newtheorem{lemma}{Lemma}

\newtheorem{example}{Example}

\newcommand{\mbf}[1]{\mathbf{#1}}

\newcommand{\kron}[2]{{#1}\otimes{#2}}

\newcommand{\F}{\mathbb{F}}

\newcommand{\vecbar}[1]{{\textrm{vec}\left\{#1\right\}}}

\begin{document}

\title{Lowest Density MDS Array Codes for Reliable Smart Meter Networks}

\author{Magnus~Sandell,~\IEEEmembership{Senior~Member,~IEEE,}
        and~Filippo~Tosato,~\IEEEmembership{Member,~IEEE}
\thanks{M. Sandell and F. Tosato are with the Telecommunications Research Laboratory of Toshiba Research Europe, 32 Queen Square, Bristol BS1 4ND, UK, email: \{magnus.sandell, filippo.tosato\}@toshiba-trel.com.}%
\thanks{This work was presented in part at the $51^{\textrm{st}}$ Annual Allerton Conference on Communication, Control and Computing.}}%

\maketitle

\begin{abstract}
In this paper we introduce a lowest density MDS array code which is applied to a Smart Meter network to introduce reliability. By treating the network as distributed storage with multiple sources, information can be exchanged between the nodes in the network allowing each node to store parity symbols relating to data from other nodes. A lowest density MDS array code is then applied to make the network robust against outages, ensuring low overhead and data transfers. We show the minimum amount of overhead required to be able to recover from $r$ node erasures in an $n$ node network and explicitly design an optimal array code with lowest density. In contrast to existing codes, this one has no restrictions on the number of nodes or erasures it can correct. Furthermore we consider incomplete networks where all nodes are not connected to each other. This limits the exchange of data for purposes of redundancy and we derive conditions on the minimum node degree that allow lowest density MDS codes to exist. We also present an explicit code design for incomplete networks that is capable of correcting two node failures.
\end{abstract}

\section{Introduction}

\begin{figure}%
\centering
{\includegraphics[width=0.6\linewidth]{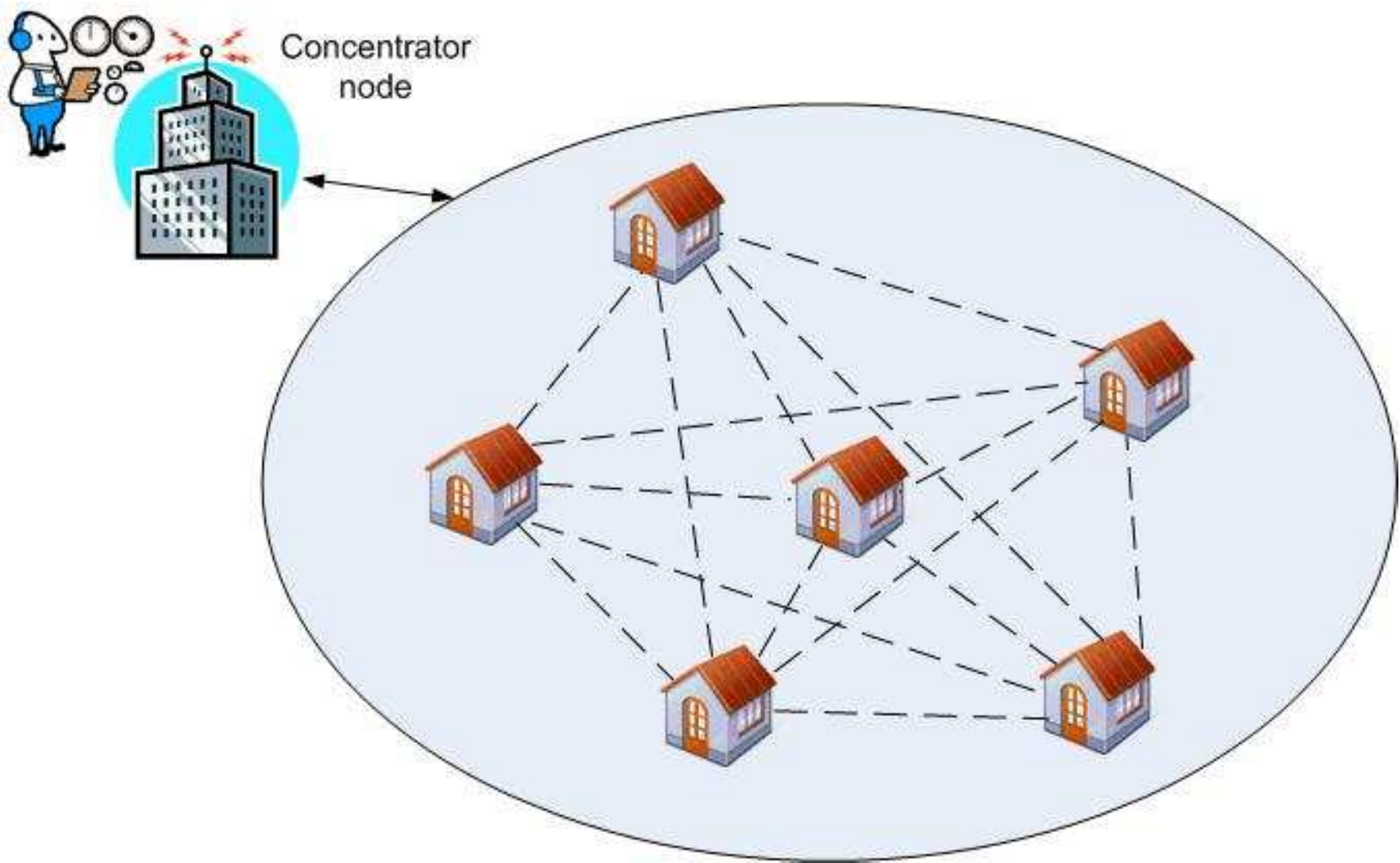}}%
\caption{Example of a Smart Meter network.}%
\label{fig:network}%
\end{figure}

It is envisaged that the modernisation of the existing energy grid will allow real-time information exchange between the utility provider and consumers to achieve more efficient energy management \cite{BAB11a}. Wireless sensors are likely to be employed for monitoring consumers' energy usage and communicating it to the utility provider, forming a Smart Meter network \cite{MaJ11c}. This is illustrated in Figure \ref{fig:network} where a number of residential and/or business properties report their consumption to a concentrator node which is located nearby. For ease of deployment, communication between the houses and/or the concentrator node is wireless, typically using existing technology such as ZigBee or IEEE802.11n. The monitoring of energy usage can be very frequent (in the order of seconds or minutes), whereas the data gathering by the concentrator node is less frequent (hours or days). This means that data is accumulated and stored at the houses (the nodes in the network) and released to the concentrator node upon request. However if all communication is wireless, nodes might only be intermittently connected to the concentrator node (at a given time, there is a chance that no reliable communication link is available). In that case, it would be desirable to obtain the whole network data by communicating with only a subset of the nodes. This is possible if redundancy is introduced in the network, \emph{i.e.}, the data from one node is stored at one or more other nodes. With an adequate scheme, it could be possible to retrieve the data from all nodes by communicating with only a subset of them; this could also be desirable even if communication with all nodes is possible. This reliability can be achieved by employing erasure coding, which makes it possible to reconstruct all the data even if some observations are erased (which is the case if some nodes are not able to communicate with the concentrator node). Maximum-distance separable (MDS) codes \cite[Ch.~11]{MacWilliams2006} are a class of erasure codes that have found extensive applications in storage systems to protect data against disk failures; the MDS property means that these codes can recover the largest possible number of erasures for a given redundancy or, equivalently, have the smallest possible overhead for a given erasure correcting capability.

Array codes are a class of MDS codes over extension alphabets that have been extensively studied in \cite{Zaitsev1981,BBB95a,Blaum1996,BlR99a,LoR06a,XuB99a,XBB99a,CaB09a}. In array codes of size $b \times n$, a single erasure affects one entire column of $b$ symbols, which can be all data symbols, all parity symbols or a combination of both. The erasure patterns that a type of MDS code can recover completely characterise the code and determine its properties. They also dictate how data and parity symbols can be separated or grouped within a codeword of length $nb$. Array codes for which any erasure deletes either all data or all parity symbols are sometimes referred to as "`horizontal"' codes \cite{LWS10a}, whereas all the other codes for which this restriction does not apply are referred to as "`vertical"'. Since each node (meter) generates its own data, the latter type is the most suitable for the Smart Meter network.

Although there are existing vertical codes, all of them have some limitation in either the number of nodes or the number of erasures they can handle. In \cite{Zaitsev1981} a family of vertical binary MDS array codes is introduced that are maximal length, \emph{i.e.}, the codeword length is the largest possible for a given block size $b$ and redundancy $r$, and have the lowest density\footnote{A formal definition of ``lowest density'' will be given in Section \ref{sec:system}.} parity check matrix. Note that, as will be discussed in Section~\ref{sec:system}, for an MDS code the redundancy equals the maximum number of erasures the code can sustain. However, the design in \cite{Zaitsev1981} is restricted to the case $r=2$, for codeword length that is a prime number and for block length $b=(n-1)/r$. In \cite{LoR06a} the construction is extended to values of $r>2$, although the resulting codes are not necessarily MDS for all parameters range. In \cite{BlR99a} a construction is presented for nonbinary systematic MDS array codes that are lowest density only for certain redundancy values. In \cite{XuB99a} a class of vertical MDS array codes is presented, called X-codes, that are lowest density; however the construction only applies if $n=b$ is prime and for $r=2$. The design is extended in \cite{XBB99a} to the so-called B-code, which is a vertical lowest density MDS array code, but the design is again restricted to $r=2$. Three other families of vertical lowest density MDS array codes are described in \cite{CaB09a} with the additional property of being cyclic or quasi-cyclic. The construction is also restricted to a specific range of parameters' values, which are summarised in Table~I of \cite{CaB09a}. Therefore all existing lowest density MDS array code constructions found in the literature have restrictions on the parameters $n$ and/or $r$. Conversely, in this work we present a code construction that works for any value of $n$ and $r$. It consists of finding the optimal size of the array, rearranging the data symbols and then applying stacked MDS codes. We will also show that this produces an MDS array code with lowest density.

\begin{figure}%
\centering
\subfloat[Complete]{\includegraphics[width=0.5\linewidth]{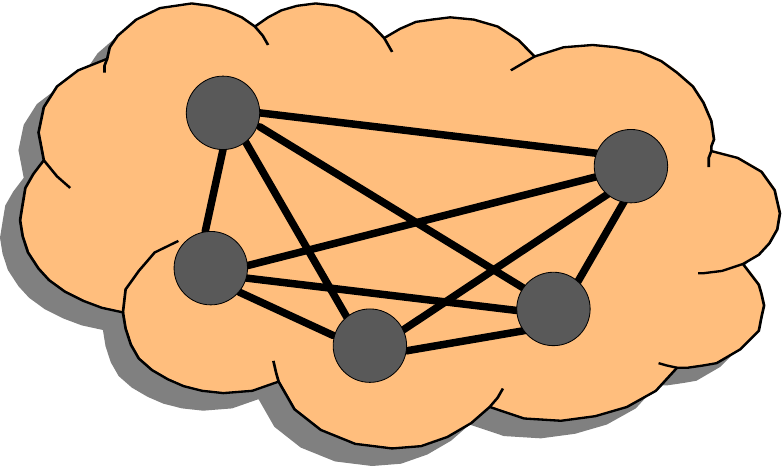}}%
\subfloat[Incomplete]{\includegraphics[width=0.47\linewidth]{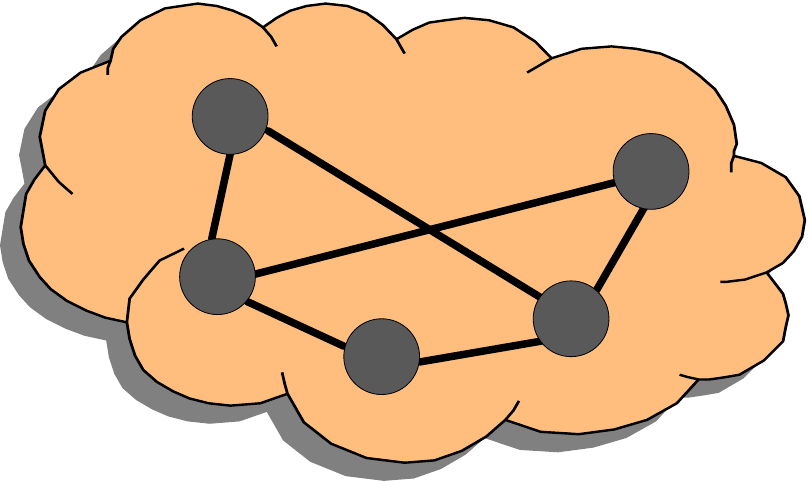}}
\caption{Two types of networks classified by their connectivity.}%
\label{fig:networks}%
\end{figure}

In general, when constructing array codes it is assumed that all nodes in the system can communicate with one another, {\emph i.e.}, all nodes can exchange information with each other. However in some applications, like the Smart Meter network studied in this paper, some nodes may not be connected to all other nodes due to topology, link failures (distance or other propagation restrictions in wireless systems) or physical properties of the network. Using terminology from graph theory, we call these two types \emph{complete} and \emph{incomplete} networks, respectively, which are illustrated in Figure \ref{fig:networks}. Although nodes could be connected via relaying or similar approaches, it might be unattractive from a traffic, latency and/or complexity point of view. Hence an erasure code must be designed such that parity symbols in a node do not depend on data symbols in nodes that are not connected to it. The previously mentioned lowest density MDS array codes do not work in this scenario as they all require all nodes to be connected to every other node. The question then arises whether lowest density MDS array codes exist for a given network with connections between only some of the nodes. We will apply graph and coding theory to study these kinds of networks and also give an explicit code design that works for all graphs with the smallest possible node degree in case of one or two node failures as well as for a subset of graphs in some other cases.

The remainder of this paper is organised as follows. In Section~\ref{sec:system} we introduce the system model for the class of codes and derive their properties, in Section~\ref{sec:design} we present the code construction and prove that the constructed codes are MDS and lowest density. Incomplete networks are treated in Section~\ref{sec:incomplete} where an explicit code design is presented for up to two node failures. Finally in Section~\ref{sec:conclusions} we draw the conclusions.

\section{System Model and Code Properties}
\label{sec:system}

\subsection{MDS property for array codes}
Let us consider a network of $n$ nodes, each generating $m$ information messages. The nodes are capable of exchanging their data messages to other nodes and storing additional $p$ parity messages in a local memory. We want to ensure that one can retrieve all $mn$ information messages in the event of up to $r$ node failures, by connecting to just $k=n-r$ nodes.

\begin{figure}%
\centering
\subfloat[Horizontal array code]{\includegraphics[width=0.4\linewidth]{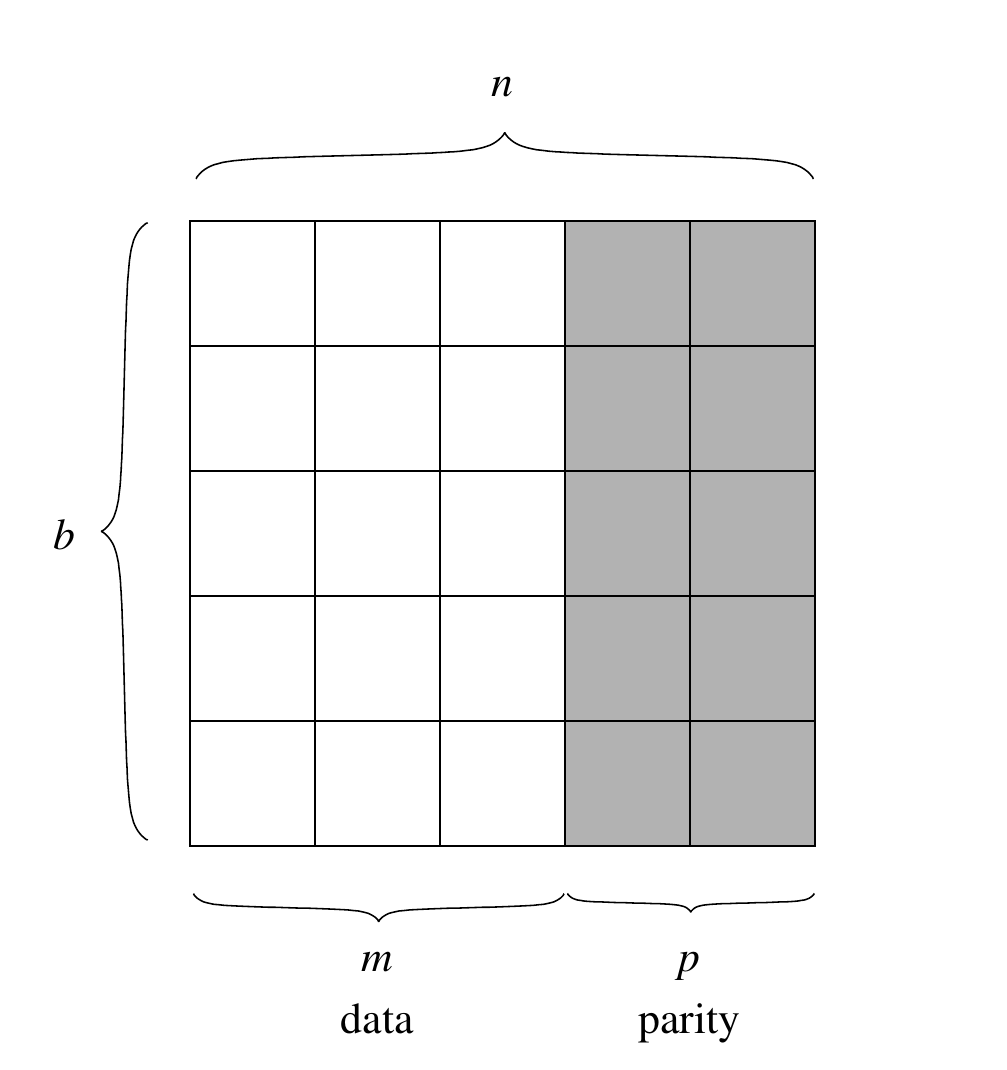}}%
\subfloat[Vertical array code]{\includegraphics[width=0.4\linewidth]{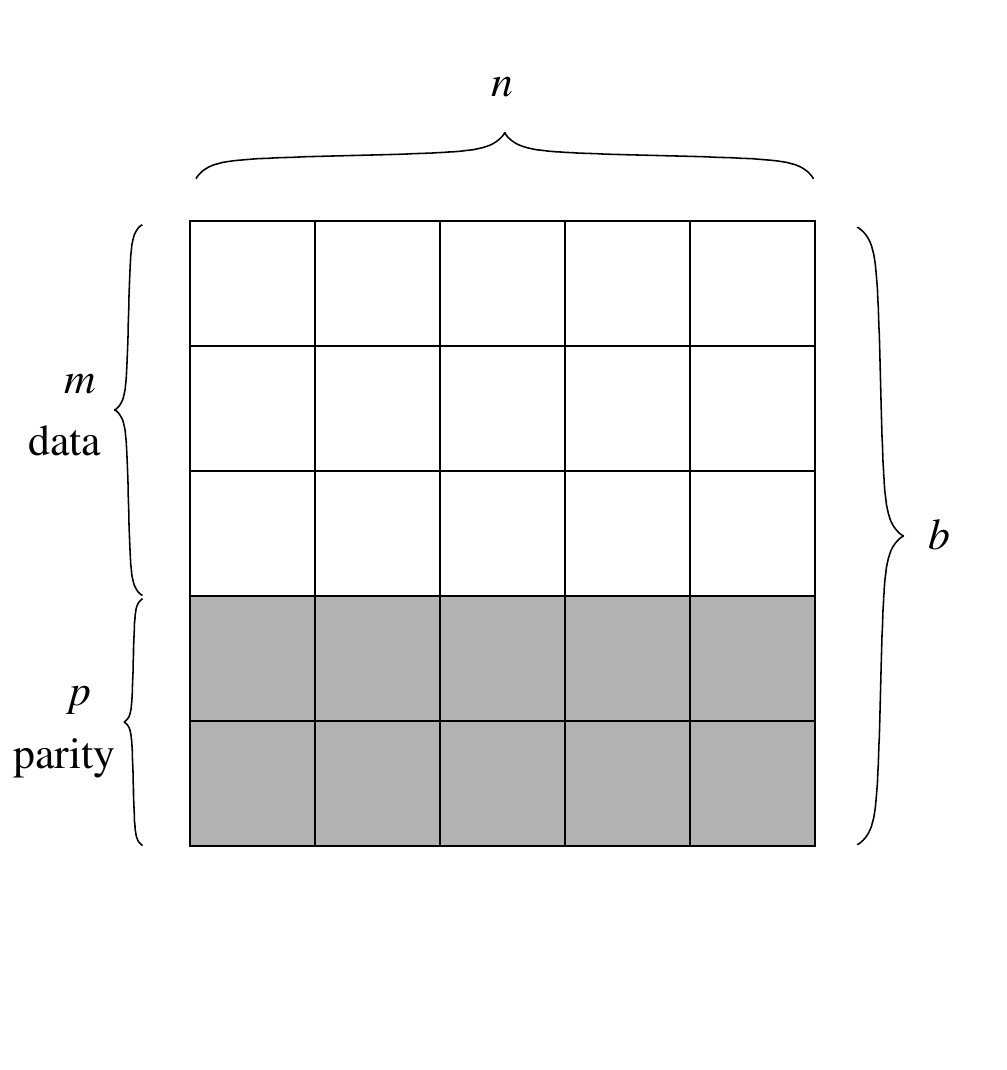}}
\caption{Two types of $b\times n$ array codes in which erasures delete entire columns.}%
\label{fig:array_codes}%
\end{figure}

This protection against node failures can be provided by erasure array codes, designed as a vertical code as illustrated in Figure \ref{fig:array_codes}. Because there are a total of $m+p$ messages stored in each node, an array code, $\mathcal{C}$, for our setup is defined by an $(m+p)\times n$ array of the form
\begin{equation}
\left(%
\begin{IEEEeqnarraybox*}[\IEEEeqnarraystrutmode\IEEEeqnarraystrutsizeadd{2pt}{0pt}][c]{x.c.c.c.x}
& d_{0,0} & \cdots & d_{0,n-1} & \\
& \vdots  & \ddots & \vdots  & \\
& d_{m-1,0} & \cdots & d_{m-1,n-1}  &\\
& f_{0,0} & \cdots & f_{0,n-1}  &\\
& \vdots  & \ddots & \vdots  &\\
& f_{p-1,0} & \cdots & f_{p-1,n-1} &
\end{IEEEeqnarraybox*}%
\right)  \label{eq:array}
\end{equation}
where $d_{i,j}$ and $f_{i,j}$ are the data and parity symbols, respectively, defined in a finite field $\F_q\triangleq\textrm{GF}(q)$. Therefore, $\mathcal{C}$ can be seen as a code of length $n$ whose ``symbols'' are $(m+p)$-blocks defined as words of the extension alphabet $\F_q^{m+p}$.

By concatenating the $(m+p)$-blocks we can represent $\mathcal{C}$ as an equivalent code of length $n(m+p)$ over $\F_q$, denoted by $\mathcal{C}_q$. If $\mathcal{C}_q$ is linear over $\F_q$, \emph{i.e.}, $\mathcal{C}_q$ forms a vector space of $\F_q$, the array code $\mathcal{C}$ is said to be an $\F_q$-linear code over $\F_q^{m+p}$. The \emph{normalised dimension} of $\mathcal{C}$, $\kappa$, is defined as the dimension of $\mathcal{C}$ measured with respect to the alphabet $\F_q^{m+p}$ and it can be calculated as the ratio between the dimension of the vector space formed by $\mathcal{C}_q$ and the column size of the array \eqref{eq:array}:
\begin{equation}
\kappa = \frac{\textrm{dim}(\mathcal{C}_q)}{m+p}=\frac{\log_{q}(|\mathcal{C}_q|)}{m+p}=\frac{nm}{m+p}.
\label{eq:kappa}
\end{equation}
Note that, in general, $\kappa$ is a rational number. Accordingly, the (normalised) redundancy of $\mathcal{C}$ is defined as $\rho=n-\kappa$, which is also, in general, a rational number. Similarly, the minimum Hamming distance, $d$, between two codewords of $\mathcal{C}$ is measured with respect to the extended alphabet $\F_q^{m+p}$ and it is always an integer number. Therefore, the Singleton bound \cite[Ch.~1,~Theorem~11]{MacWilliams2006} dictates that for an $[n,\kappa, d]$ $\F_q$-linear array code over $\F_q^{m+p}$ the following inequality holds
\begin{equation}
d\leq n-\kappa+1=n-\frac{nm}{m+p}+1.
\label{eq:singleton_bound}
\end{equation}
Because we are interested in such codes that are MDS, \emph{i.e.}, they offer maximum erasure correction capability with minimum redundancy, we require that (\ref{eq:singleton_bound}) holds with equality. In this case we can prove the following result

\begin{proposition}\label{prop:pk_mr}
An $[n,\kappa,d]$ MDS array code over $\F_q^{m+p}$, $\mathcal{C}$, capable of correcting up to $r=n-k$ node failures must satisfy the following relationship:
\begin{equation}
pk = mr
\label{eq:MDS_necessary}
\end{equation}
\end{proposition}
\begin{IEEEproof}
Because an array code is able to correct up to $d-1$ block erasures, we have $d-1=r$. From the MDS property, by taking (\ref{eq:singleton_bound}) with the equality sign we obtain
\begin{equation}
r+1 = d = n-\frac{nm}{m+p}+1
\end{equation}
and hence, by using $r=n-k$,
\begin{equation}
k=\frac{nm}{m+p}
\label{eq:k_MDS}
\end{equation}
from which (\ref{eq:MDS_necessary}) follows.
\end{IEEEproof}
By comparing (\ref{eq:kappa}) and (\ref{eq:k_MDS}) it follows that the normalised dimension of an MDS array code is an integer number and $\kappa=k$. Consequently, the normalised redundancy is $\rho=r$. Henceforth we will use the parameters pair $[n,k]$ to denote an MDS array code.

Note that (\ref{eq:MDS_necessary}) is a necessary but not sufficient condition to characterise this class of MDS array codes. In fact, different types of MDS codes may be defined by different patterns of erasures. In a generic $[n,k]$ linear MDS code over $\F_q$, any of the $n\choose r$ combinations of $r=n-k$ erasures are admissible, hence any $k$ symbols of a codeword can be used to reconstruct the message. This property turns out to be necessary and sufficient to characterise an MDS code over $\F_q$ \cite[Ch.~11, Corollary~3]{MacWilliams2006}. On the other hand, for MDS array codes, a single erasure consists in the loss of a number of symbols of $\F_q$, which can be a combination of systematic and/or parity symbols depending on the code definition. In this paper, because of the nature of the problem formulation, we deal with vertical MDS array codes in which a symbol of the extension alphabet $\F_q^{m+p}$ contains exactly $m$ data and $p$ parity messages stored in a single node.

\subsection{Generator matrix}

The generator and parity matrices of an $[n,k]$, $\F_q$-linear array code over $\F_q^{m+p}$, $\mathcal{C}$, are defined as the generator and parity matrices, respectively, of its linearised version $\mathcal{C}_q$. It is useful for the purpose of constructing the code to introduce a representation of the generator matrix of $\mathcal{C}$ in which the columns are arranged in systematic form. We stress that arranging the generator matrix in a systematic form does not affect the properties of the code which are dictated by the patterns of erasures that a code can correct. The generator matrix reads
\begin{equation}
\mbf{G}=\left(\begin{IEEEeqnarraybox*}[\IEEEeqnarraystrutmode\IEEEeqnarraystrutsizeadd{2pt}{0pt}][c]{x.c.vv.c.x} & \mbf{I}_{nm} &&& \mbf{A} &\end{IEEEeqnarraybox*}\right)
\label{eq:G_notation1}
\end{equation}
where $\mbf{A}$, of size $nm\times np$, denotes the nonsystematic part. The array code encodes the data message as a vector-matrix multiplication, where $\mbf{c}$ is the codeword of length $n(m+p)$ and $\mbf{d}$ is the array of data symbols of length $nm$
\begin{align}
\mbf{c} &=(\begin{IEEEeqnarraybox*}[\IEEEeqnarraystrutmode\IEEEeqnarraystrutsizeadd{2pt}{0pt}][c]{x.c.c.c.c.c.c.c.c.c.c.c.c.x}
& d_{0,0} & \cdots & d_{m-1,0} & d_{0,1} & \cdots & d_{m-1,n-1} & f_{0,0} & \cdots & f_{p-1,0} & f_{0,1} & \cdots & f_{p-1,n-1} &
\end{IEEEeqnarraybox*}) \label{eq:c_notation1}\\
&=(\begin{IEEEeqnarraybox*}[\IEEEeqnarraystrutmode\IEEEeqnarraystrutsizeadd{2pt}{0pt}][c]{x.c.c.c.c.c.c.x}
& d_{0,0} & \cdots & d_{m-1,0} & d_{0,1} & \cdots & d_{m-1,n-1} &\end{IEEEeqnarraybox*})\left(%
\begin{IEEEeqnarraybox*}[\IEEEeqnarraystrutmode\IEEEeqnarraystrutsizeadd{2pt}{0pt}][c]{x.c.c.c.c.vv.c.c.c.x}
& \mathbf{I}_{m} & \mathbf{0} & \cdots & \mathbf{0} &&& \mathbf{A}_{0,0} & \cdots & \mathbf{A}_{0,n-1} & \\
& \mathbf{0} & \mathbf{I}_{m} & \cdots & \mathbf{0} &&& \mathbf{A}_{1,0} & \cdots & \mathbf{A}_{1,n-1} & \\
& \vdots & \vdots & \ddots & \vdots &&& \vdots & \ddots & \vdots & \\
& \mathbf{0} & \cdots & \cdots & \mathbf{I}_{m} &&& \mathbf{A}_{n-1,0} & \cdots & \mathbf{A}_{n-1,n-1} &
\end{IEEEeqnarraybox*}
\right) \label{eq:block_A}\\
&= \mbf{d}\mbf{G}.
\label{eq:dG}
\end{align}

In the representation in \eqref{eq:block_A} the $\mbf{A}$ matrix can be partitioned in $n\times n$ blocks of size $m\times p$: $\mbf{A}=[\mbf{A}_{i,j}]_{i,j=0}^{n-1}$. In the event of $r=n-k$ node failures, let $\mathcal{F}=\{i_0,\dots,i_{r-1}\}$ be the set of failing nodes and $\mathcal{S}=\{i_r,\dots,i_{n-1}\}=\{0,\dots,n-1\}\setminus\mathcal{F}$ the set of surviving nodes. The $r$ columns of (\ref{eq:array}) indexed by $\mathcal{F}$ become unavailable and $r(m+p)$ values of (\ref{eq:c_notation1}) are erased. This is the same as removing $2r$ block columns from (\ref{eq:G_notation1}): $r$ block columns from $\mbf{I}_{nm}$ and $r$ block columns from $\mbf{A}$. From the surviving nodes we can directly retrieve the following $km$ data symbols
\begin{equation}
\mbf{d}^{(s)} = (\begin{IEEEeqnarraybox*}[\IEEEeqnarraystrutmode\IEEEeqnarraystrutsizeadd{2pt}{0pt}][c]{x.c.c.c.c.c.c.x}& d_{0,i_r} & \cdots & d_{m-1,i_r} & d_{0,i_{r+1}} & \cdots & d_{m-1,i_{n-1}}\end{IEEEeqnarraybox*})
\end{equation}
whilst the remaining $rm$ messages, originated in the failed nodes, can be grouped in the vector
\begin{equation}
\mbf{d}^{(f)} = (
\begin{IEEEeqnarraybox*}[\IEEEeqnarraystrutmode\IEEEeqnarraystrutsizeadd{2pt}{0pt}][c]{x.c.c.c.c.c.c.x}
& d_{0,i_0} & \cdots & d_{m-1,i_0} & d_{0,i_1} & \cdots & d_{m-1,i_{r-1}}
\end{IEEEeqnarraybox*}).
\label{eq:df}
\end{equation}
Accordingly, we partition the $k$ block columns of $\mbf{A}$ indexed by $\mathcal{S}$ in two submatrices
\begin{align}
\mbf{A}^{(s)} &=[\mbf{A}_{i,j}]_{i,j\in\mathcal{S}} \in \F_q^{km \times kp} \label{eq:As}\\
\mbf{A}^{(f)} &=[\mbf{A}_{i,j}]_{i\in\mathcal{F},j\in\mathcal{S}} \in \F_q^{rm \times kp} \label{eq:Af}
\end{align}
representing the contribution to the surviving parity symbols of the surviving nodes and failed nodes, respectively. Note that the rows and columns of $\mbf{A}$ forming $\mbf{A}^{(f)}$ are mutually exclusive since a node either fails or survives. Therefore, we can expand the surviving parity symbol equation as follows
\begin{align}
\mbf{f}^{(s)} &= (\begin{IEEEeqnarraybox*}[\IEEEeqnarraystrutmode\IEEEeqnarraystrutsizeadd{2pt}{0pt}][c]{x.c.c.c.c.c.c.x}& f_{0,i_r} & \cdots & f_{p-1,i_r} & f_{0,i_{r+1}} & \cdots & f_{p-1,i_{n-1}}\end{IEEEeqnarraybox*})\\
&= \mbf{d}^{(s)}\mbf{A}^{(s)} + \mbf{d}^{(f)}\mbf{A}^{(f)}
\end{align}
and solve for the unknowns $\mbf{d}^{(f)}$ to recover the $rm$ data messages of the failed nodes:
\begin{equation}
\mbf{d}^{(f)} = (\mbf{f}^{(s)} - \mbf{d}^{(s)}\mbf{A}^{(s)}){\mbf{A}^{(f)}}^{-1}.
\label{eq:decoding}
\end{equation}

In order for (\ref{eq:decoding}) to have a unique solution, any matrix $\mbf{A}^{(f)}$ must have rank $rm$. In fact, this is a necessary and sufficient condition to characterise our class of MDS codes:

\begin{proposition}\label{prop:NecAndSuf}
An $[n,k=n-r]$, $\F_q$-linear array code over $\F_q^{m+p}$, $\mathcal{C}$, is MDS iff all the $n\choose r$ submatrices $\mbf{A}^{(f)}$ are nonsingular.
\end{proposition}
\begin{IEEEproof}
Necessity: if $\mathcal{C}$ is MDS, from Proposition~\ref{prop:pk_mr} the code can correct any $r$ node failures and the $\mbf{A}^{(f)}$ matrices are square. It follows immediately from (\ref{eq:decoding}) that the $\mbf{A}^{(f)}$ matrices are nonsingular.

Sufficiency: if all the $\mbf{A}^{(f)}$ matrices are nonsingular, the code can correct any $r$ node failures as shown in the derivation of (\ref{eq:decoding}), hence its minimum distance, $d$, must be at least $r+1$. Because $r$ is also the code redundancy, by the Singleton bound (\ref{eq:singleton_bound}) it follows that $d=r+1$, hence $\mathcal{C}$ is MDS.
\end{IEEEproof}

Note that the characterisation of Proposition~\ref{prop:NecAndSuf} is different from \cite[Ch.~11, Theorem~8]{MacWilliams2006}, which applies to MDS linear codes over $\F_q$ and from \cite[Proposition~3.2]{BlR99a}, which applies to horizontal MDS array codes. This different characterisation stems from the different types of error patterns that our class of codes has to correct.

\subsection{Duality}
All MDS array codes have a dual code and we can prove the following result on their relationship.

\begin{proposition}[Duality]\label{prop:duality}
The dual code, $\mathcal{C}^{\perp}$, of an $[n,k=n-r]$, $\F_q$-linear MDS array code over $\F_q^{m+p}$, $\mathcal{C}$, with $m$ data and $p$ parity messages per node, is an $[n,r]$, $\F_q$-linear MDS array code over $\F_q^{m+p}$ with $p$ data and $m$ parity messages per node.
\end{proposition}
\begin{IEEEproof}
The parity matrix of $\mathcal{C}$, $\mbf{H}=(-\mbf{A}^T\,|\,\mbf{I}_{np})$, is a generator matrix of $\mathcal{C}^\perp$. It is not difficult to see that if Proposition~\ref{prop:NecAndSuf} holds true for a code having $\mbf{A}$ as nonsystematic part of the generator matrix, then the same must be true for $\mathcal{C}^\perp$, whose nonsystematic part is given by $-\mbf{A}^T$, once $k$ is exchanged with $r$ and $m$ with $p$.
\end{IEEEproof}

Because of this duality principle, one may restrict the analysis, without loss of generality, to the case $k\geq r$. Moreover, the smallest possible choice of the parameters $m$ and $p$ is to make them coprime by choosing them as
\begin{subequations}
\label{eq:m-p-MDS}
\begin{align}
m & = \frac{k}{\gcd ( k,r )} \\
p & = \frac{r}{\gcd ( k,r )}.
\end{align}
\end{subequations}
This choice ensures the smallest possible array. Once an $[n,k]$ MDS array code is available for a coprime pair $m,p$, an extended MDS code can be easily constructed for the pair $am,ap$, for any positive integer $a$. This extended family of codes are found by simply reusing the same mother code $a$ times. It can be easily seen that the generator matrix of the extended code, $\mbf{G}_a$, is obtained by a Kronecker product
\begin{equation}
\mbf{G}_a=\kron{\mbf{G}}{\mbf{I}_a}.
\label{eq:kron_extension}
\end{equation}

\subsection{Lowest density}
It is also desirable for an MDS code to have the smallest possible number of nonzero entries in the generator matrix, \emph{i.e.}, a very low-density parity check (LDPC) matrix. The update complexity is the number of parity check symbols that need to be modified every time a data symbol is updated. For message $i=0,\dots,m-1$ of node $j=0,\dots,n-1$, an update of data symbol $d_{i,j}$ requires modification to a number of parity symbols equal to the Hamming weight of row $i+jm$ of $\mbf{A}$. This is because the number of parity symbols that a data symbol appears in is given by the weight of the corresponding row of $\mbf{A}$. Therefore, minimising the update complexity is achieved by minimising the number of nonzero entries in the $\mbf{A}$ matrix. To this end, we can prove the following result.

\begin{proposition}\label{prop:lowest_density}
For an $[n,k=n-r]$, MDS array code over $\F_q^{m+p}$, the minimum row and column weight of the nonsystematic part of the generator matrix, $\mbf{A}$, is $r$ and $k$, respectively. Moreover, if each row has exactly weight $r$, then each column must have weight $k$ and vice versa.
\end{proposition}
\begin{IEEEproof}
By definition each row of the generator matrix (\ref{eq:G_notation1}) is a codeword, obtained from (\ref{eq:dG}) by replacing $\mbf{d}$ with the appropriate unit vector. Therefore, because the minimum distance of the code is $r+1$, and one nonzero element in each row of $\mbf{G}$ is in the systematic part, the minimum row weight of $\mbf{A}$ is $r$. From Proposition~\ref{prop:duality}, the dual code is also MDS with parameters $k$ and $r$ exchanged, hence the minimum row weight of $\mbf{A}^T$ must be $k$, which proves the first part of the assert.

With regard to the second statement, if each row of $\mbf{A}$ has weight $r$, then there are in total $rnm$ nonzero elements in $\mbf{A}$. Hence, the average number of nonzero elements per column is $rnm/np=k$, where we used the property (\ref{eq:MDS_necessary}). However, because $k$ is also the minimum column weight, each column must have exactly $k$ nonzero elements. Vice versa is also true by the code duality argument.
\end{IEEEproof}

A similar result to Proposition~\ref{prop:lowest_density} was proved in \cite{BlR99a} for a different class of MDS array codes. We refer to an MDS array code of the type defined in this paper as \emph{lowest density} when each row of $\mbf{A}$ contains exactly $r$ nonzero elements.

\section{Code design}
\label{sec:design}

In this section we will show how a vertical MDS array code can be designed for any number of nodes and any number of erasures. If rowwise encoding of the array in \eqref{eq:array} is used, $r$ nodes will consist of only parity symbols (it will be a horizontal code) and consequently it will not be applicable to the Smart Meter network. However it can be made into a vertical code by arranging the data symbols in the array in an appropriate manner, which is described in Algorithm \ref{alg:array_code}; this procedure forms an MDS array code. Note that this arrangement of data symbols in the array is not unique but can be done in several equivalent ways.
\begin{proposition}
\label{prop:alg-is-MDS}
The array code in Algorithm \ref{alg:array_code} is a $[n,k]$ MDS code capable of correcting $r$ column erasures.
\end{proposition}
\begin{IEEEproof}
First we show that all the rows of the array have $k$ data symbols and $n-k$ parity symbols. Since $0 \leq i < m,~0 \leq j < n$, the term $i+jm$ will enumerate all integers from $0$ to $nm-1$ and as we have that $nm \mod (m+p) = 0$ from \eqref{eq:k_MDS}, the row index will have $\frac{nm}{m+p}=k$ occurrences of each integer. This mean that there will be exactly $k$ data symbols per row and consequently $n-k=r$ empty slots. These can be filled with parity symbols from a $[n,k]$ systematic linear MDS code. Since this code is capable of correcting $r$ erasures, the array is able to recover from any $r$ erased columns.
\end{IEEEproof}

It is worth noting that since the array is encoded rowwise, each row can be decoded separately. As each individual code is only a one-dimensional $[n,k]$ code, decoding complexity can be kept low and also facilitate parallelism.

\begin{proposition}
\label{prop:stacked_code}
The generator matrix for the code in Algorithm \ref{alg:array_code} can be expressed as a Kronecker product
\begin{align}
\mbf{A} &= \tilde{\mbf{A}} \otimes \mbf{D} \in \F_q ^{nm \times np} \label{eq:our-code} \\
\tilde{\mbf{A}} & \in \F_q^{k \times r}, \text{totally nonsingular\footnotemark} \\
\mbf{D} & \in \F_q^{(m+p) \times (m+p)} \\
D_{i,j} &= \left\{
\begin{aligned}
1, & \hspace*{5mm} i+j+1 \mod (m+p) = 0 \\
0, & \hspace*{5mm} \text{otherwise}
\end{aligned}
\right., i,j=0,\cdots,m+p-1
\end{align}
\end{proposition}
\footnotetext{All square submatrices are nonsingular.}

\begin{IEEEproof}
See Appendix \ref{app:stacked_code}.
\end{IEEEproof}

The code design can be illustrated with an example.

\begin{example}
\label{ex:array}
Consider a network with $n=8$ nodes capable of correcting $r=3$ erased nodes. Choosing the number of data and parity symbols per nodes as $m = 5/\gcd(5,3) = 5,~p = 3 \cdot 5/5 = 3$, we can put data symbol $d_{i,j}$ in row $(i+5j) \mod 8$, column $j$ and parity symbol $f_{i,j}$ in row $(5j-i-1) \mod 8$, column $j$. This creates the array
\begin{equation}
\left(
\begin{array}{cccccccc}
d_{0,0} & d_{3,1} & f_{1,2} & d_{1,3} & d_{4,4} & f_{0,5} & d_{2,6} & f_{2,7} \\
d_{1,0} & d_{4,1} & f_{0,2} & d_{2,3} & f_{2,4} & d_{0,5} & d_{3,6} & f_{1,7} \\
d_{2,0} & f_{2,1} & d_{0,2} & d_{3,3} & f_{1,4} & d_{1,5} & d_{4.6} & f_{0,7} \\
d_{3,0} & f_{1,1} & d_{1,2} & d_{4,3} & f_{0,4} & d_{2,5} & f_{2,6} & d_{0,7} \\
d_{4,0} & f_{0,1} & d_{2,2} & f_{2,3} & d_{0,4} & d_{3,5} & f_{1,6} & d_{1,7} \\
f_{2,0} & d_{0,1} & d_{3,2} & f_{1,3} & d_{1,4} & d_{4,5} & f_{0,6} & d_{2,7} \\
f_{1,0} & d_{1,1} & d_{4,2} & f_{0,3} & d_{2,4} & f_{2,5} & d_{0,6} & d_{3,7} \\
f_{0,0} & d_{2,1} & f_{2,2} & d_{0,3} & d_{3,4} & f_{1,5} & d_{1,6} & d_{4,7}
\end{array}
\right).
\label{eq:array-example}
\end{equation}
It is clear that there are $r=3$ parity symbols in each row which can be calculated from a $[8,5]$ systematic MDS code. It can be created as a $q$-ary generalised Reed-Solomon code by using a Singleton matrix \cite{RoS85a} which is totally nonsingular. The rowwise generator matrix could then be
\begin{eqnarray}
\tilde{\mbf{G}} &=& \left(\begin{IEEEeqnarraybox*}[\IEEEeqnarraystrutmode\IEEEeqnarraystrutsizeadd{2pt}{0pt}][c]{x.c.vv.c.x} & \mbf{I}_{5} &&& \tilde{\mbf{A}} &\end{IEEEeqnarraybox*}\right) \in \F_{7}^{5 \times 8} \\
\tilde{\mbf{A}} &=& \left(
\begin{array}{ccc}
1 & 1 & 1 \\
1 & 3 & 6 \\
1 & 4 & 2 \\
1 & 6 & 4 \\
1 & 2 & 5
\end{array}
\right) \in \F_{7}^{5 \times 3}.
\label{eq:RS-matrix}
\end{eqnarray}
The parity symbols in the array are then formed as
\begin{equation}
\label{eq:rowwise}
\left(
\begin{array}{ccc}
f_{1,2} & f_{0,5} & f_{2,7} \\
f_{0,2} & f_{2,4} & f_{1,7} \\
f_{2,1} & f_{1,4} & f_{0,7} \\
f_{1,1} & f_{0,4} & f_{2,6} \\
f_{0,1} & f_{2,3} & f_{1,6} \\
f_{2,0} & f_{1,3} & f_{0,6} \\
f_{1,0} & f_{0,3} & f_{2,5} \\
f_{0,0} & f_{2,2} & f_{1,5}
\end{array}
\right) = \left(
\begin{array}{ccccc}
d_{0,0} & d_{3,1} & d_{1,3} & d_{4,4} & d_{2,6} \\
d_{1,0} & d_{4,1} & d_{2,3} & d_{0,5} & d_{3,6} \\
d_{2,0} & d_{0,2} & d_{3,3} & d_{1,5} & d_{4,6} \\
d_{3,0} & d_{1,2} & d_{4,3} & d_{2,5} & d_{0,7} \\
d_{4,0} & d_{2,2} & d_{0,4} & d_{3,5} & d_{1,7} \\
d_{0,1} & d_{3,2} & d_{1,4} & d_{4,5} & d_{2,7} \\
d_{1,1} & d_{4,2} & d_{2,4} & d_{0,6} & d_{3,7} \\
d_{2,1} & d_{0,3} & d_{3,4} & d_{1,6} & d_{4,7}
\end{array}
\right) \tilde{\mbf{A}}.
\end{equation}
Since each row is capable of correcting three erasures, the whole array can recover from any three erased columns. Note that although the vertical codes in \cite{LoR06a,LWS10a,CaB09a} are capable of correcting three erasures, none of them will work for $n=8$ nodes. Hence the design described above, which works for any number of nodes and erasures, can be designed for any network configuration and is not limited to special cases.

The explicit encoding of the parity symbols is now
\begin{align}
\mbf{f} &=(\begin{IEEEeqnarraybox*}[\IEEEeqnarraystrutmode\IEEEeqnarraystrutsizeadd{2pt}{0pt}][c]{x.c.c.c.c.c.c.x}
& f_{0,0} & f_{1,0} & f_{2,0} & f_{0,1} &  \cdots & f_{2,7} &
\end{IEEEeqnarraybox*}) \\
 &=(\begin{IEEEeqnarraybox*}[\IEEEeqnarraystrutmode\IEEEeqnarraystrutsizeadd{2pt}{0pt}][c]{x.c.c.c.c.c.c.c.c.x}
& d_{0,0} & d_{1,0} & d_{2,0} & d_{3,0} & d_{4,0} & d_{0,1} & \cdots & d_{4,7} &
\end{IEEEeqnarraybox*})  \times  \nonumber \\
& \left(%
\begin{IEEEeqnarraybox*}[\IEEEeqnarraystrutmode\IEEEeqnarraystrutsizeadd{-5pt}{-5pt}][c]{x.c.c.c.c.c.c.c.c.c.c.c.c.c.c.c.c.c.c.c.c.c.c.c.c.x}
& 0 & 0 & 0 & 0 & 0 & 0 & 0 & 1 & 0 & 0 & 0 & 0 & 0 & 0 & 0 & 1 & 0 & 0 & 0 & 0 & 0 & 0 & 0 & 1 & \\
& 0 & 0 & 0 & 0 & 0 & 0 & 1 & 0 & 0 & 0 & 0 & 0 & 0 & 0 & 1 & 0 & 0 & 0 & 0 & 0 & 0 & 0 & 1 & 0 & \\
& 0 & 0 & 0 & 0 & 0 & 1 & 0 & 0 & 0 & 0 & 0 & 0 & 0 & 1 & 0 & 0 & 0 & 0 & 0 & 0 & 0 & 1 & 0 & 0 & \\
& 0 & 0 & 0 & 0 & 1 & 0 & 0 & 0 & 0 & 0 & 0 & 0 & 1 & 0 & 0 & 0 & 0 & 0 & 0 & 0 & 1 & 0 & 0 & 0 & \\
& 0 & 0 & 0 & 1 & 0 & 0 & 0 & 0 & 0 & 0 & 0 & 1 & 0 & 0 & 0 & 0 & 0 & 0 & 0 & 1 & 0 & 0 & 0 & 0 & \\
& 0 & 0 & 1 & 0 & 0 & 0 & 0 & 0 & 0 & 0 & 1 & 0 & 0 & 0 & 0 & 0 & 0 & 0 & 1 & 0 & 0 & 0 & 0 & 0 & \\
& 0 & 1 & 0 & 0 & 0 & 0 & 0 & 0 & 0 & 1 & 0 & 0 & 0 & 0 & 0 & 0 & 0 & 1 & 0 & 0 & 0 & 0 & 0 & 0 & \\
& 1 & 0 & 0 & 0 & 0 & 0 & 0 & 0 & 1 & 0 & 0 & 0 & 0 & 0 & 0 & 0 & 1 & 0 & 0 & 0 & 0 & 0 & 0 & 0 & \\
& 0 & 0 & 0 & 0 & 0 & 0 & 0 & 1 & 0 & 0 & 0 & 0 & 0 & 0 & 0 & 3 & 0 & 0 & 0 & 0 & 0 & 0 & 0 & 6 & \\
& 0 & 0 & 0 & 0 & 0 & 0 & 1 & 0 & 0 & 0 & 0 & 0 & 0 & 0 & 3 & 0 & 0 & 0 & 0 & 0 & 0 & 0 & 6 & 0 & \\
& 0 & 0 & 0 & 0 & 0 & 1 & 0 & 0 & 0 & 0 & 0 & 0 & 0 & 3 & 0 & 0 & 0 & 0 & 0 & 0 & 0 & 6 & 0 & 0 & \\
& 0 & 0 & 0 & 0 & 1 & 0 & 0 & 0 & 0 & 0 & 0 & 0 & 3 & 0 & 0 & 0 & 0 & 0 & 0 & 0 & 6 & 0 & 0 & 0 & \\
& 0 & 0 & 0 & 1 & 0 & 0 & 0 & 0 & 0 & 0 & 0 & 3 & 0 & 0 & 0 & 0 & 0 & 0 & 0 & 6 & 0 & 0 & 0 & 0 & \\
& 0 & 0 & 1 & 0 & 0 & 0 & 0 & 0 & 0 & 0 & 3 & 0 & 0 & 0 & 0 & 0 & 0 & 0 & 6 & 0 & 0 & 0 & 0 & 0 & \\
& 0 & 1 & 0 & 0 & 0 & 0 & 0 & 0 & 0 & 3 & 0 & 0 & 0 & 0 & 0 & 0 & 0 & 6 & 0 & 0 & 0 & 0 & 0 & 0 & \\
& 1 & 0 & 0 & 0 & 0 & 0 & 0 & 0 & 3 & 0 & 0 & 0 & 0 & 0 & 0 & 0 & 6 & 0 & 0 & 0 & 0 & 0 & 0 & 0 & \\
& 0 & 0 & 0 & 0 & 0 & 0 & 0 & 1 & 0 & 0 & 0 & 0 & 0 & 0 & 0 & 4 & 0 & 0 & 0 & 0 & 0 & 0 & 0 & 2 & \\
& 0 & 0 & 0 & 0 & 0 & 0 & 1 & 0 & 0 & 0 & 0 & 0 & 0 & 0 & 4 & 0 & 0 & 0 & 0 & 0 & 0 & 0 & 2 & 0 & \\
& 0 & 0 & 0 & 0 & 0 & 1 & 0 & 0 & 0 & 0 & 0 & 0 & 0 & 4 & 0 & 0 & 0 & 0 & 0 & 0 & 0 & 2 & 0 & 0 & \\
& 0 & 0 & 0 & 0 & 1 & 0 & 0 & 0 & 0 & 0 & 0 & 0 & 4 & 0 & 0 & 0 & 0 & 0 & 0 & 0 & 2 & 0 & 0 & 0 & \\
& 0 & 0 & 0 & 1 & 0 & 0 & 0 & 0 & 0 & 0 & 0 & 4 & 0 & 0 & 0 & 0 & 0 & 0 & 0 & 2 & 0 & 0 & 0 & 0 & \\
& 0 & 0 & 1 & 0 & 0 & 0 & 0 & 0 & 0 & 0 & 4 & 0 & 0 & 0 & 0 & 0 & 0 & 0 & 2 & 0 & 0 & 0 & 0 & 0 & \\
& 0 & 1 & 0 & 0 & 0 & 0 & 0 & 0 & 0 & 4 & 0 & 0 & 0 & 0 & 0 & 0 & 0 & 2 & 0 & 0 & 0 & 0 & 0 & 0 & \\
& 1 & 0 & 0 & 0 & 0 & 0 & 0 & 0 & 4 & 0 & 0 & 0 & 0 & 0 & 0 & 0 & 2 & 0 & 0 & 0 & 0 & 0 & 0 & 0 & \\
& 0 & 0 & 0 & 0 & 0 & 0 & 0 & 1 & 0 & 0 & 0 & 0 & 0 & 0 & 0 & 6 & 0 & 0 & 0 & 0 & 0 & 0 & 0 & 4 & \\
& 0 & 0 & 0 & 0 & 0 & 0 & 1 & 0 & 0 & 0 & 0 & 0 & 0 & 0 & 6 & 0 & 0 & 0 & 0 & 0 & 0 & 0 & 4 & 0 & \\
& 0 & 0 & 0 & 0 & 0 & 1 & 0 & 0 & 0 & 0 & 0 & 0 & 0 & 6 & 0 & 0 & 0 & 0 & 0 & 0 & 0 & 4 & 0 & 0 & \\
& 0 & 0 & 0 & 0 & 1 & 0 & 0 & 0 & 0 & 0 & 0 & 0 & 6 & 0 & 0 & 0 & 0 & 0 & 0 & 0 & 4 & 0 & 0 & 0 & \\
& 0 & 0 & 0 & 1 & 0 & 0 & 0 & 0 & 0 & 0 & 0 & 6 & 0 & 0 & 0 & 0 & 0 & 0 & 0 & 4 & 0 & 0 & 0 & 0 & \\
& 0 & 0 & 1 & 0 & 0 & 0 & 0 & 0 & 0 & 0 & 6 & 0 & 0 & 0 & 0 & 0 & 0 & 0 & 4 & 0 & 0 & 0 & 0 & 0 & \\
& 0 & 1 & 0 & 0 & 0 & 0 & 0 & 0 & 0 & 6 & 0 & 0 & 0 & 0 & 0 & 0 & 0 & 4 & 0 & 0 & 0 & 0 & 0 & 0 & \\
& 1 & 0 & 0 & 0 & 0 & 0 & 0 & 0 & 6 & 0 & 0 & 0 & 0 & 0 & 0 & 0 & 4 & 0 & 0 & 0 & 0 & 0 & 0 & 0 & \\
& 0 & 0 & 0 & 0 & 0 & 0 & 0 & 1 & 0 & 0 & 0 & 0 & 0 & 0 & 0 & 2 & 0 & 0 & 0 & 0 & 0 & 0 & 0 & 5 & \\
& 0 & 0 & 0 & 0 & 0 & 0 & 1 & 0 & 0 & 0 & 0 & 0 & 0 & 0 & 2 & 0 & 0 & 0 & 0 & 0 & 0 & 0 & 5 & 0 & \\
& 0 & 0 & 0 & 0 & 0 & 1 & 0 & 0 & 0 & 0 & 0 & 0 & 0 & 2 & 0 & 0 & 0 & 0 & 0 & 0 & 0 & 5 & 0 & 0 & \\
& 0 & 0 & 0 & 0 & 1 & 0 & 0 & 0 & 0 & 0 & 0 & 0 & 2 & 0 & 0 & 0 & 0 & 0 & 0 & 0 & 5 & 0 & 0 & 0 & \\
& 0 & 0 & 0 & 1 & 0 & 0 & 0 & 0 & 0 & 0 & 0 & 2 & 0 & 0 & 0 & 0 & 0 & 0 & 0 & 5 & 0 & 0 & 0 & 0 & \\
& 0 & 0 & 1 & 0 & 0 & 0 & 0 & 0 & 0 & 0 & 2 & 0 & 0 & 0 & 0 & 0 & 0 & 0 & 5 & 0 & 0 & 0 & 0 & 0 & \\
& 0 & 1 & 0 & 0 & 0 & 0 & 0 & 0 & 0 & 2 & 0 & 0 & 0 & 0 & 0 & 0 & 0 & 5 & 0 & 0 & 0 & 0 & 0 & 0 & \\
& 1 & 0 & 0 & 0 & 0 & 0 & 0 & 0 & 2 & 0 & 0 & 0 & 0 & 0 & 0 & 0 & 5 & 0 & 0 & 0 & 0 & 0 & 0 & 0 & \\
\end{IEEEeqnarraybox*}
\right) \\
&= \left( \begin{array}{ccc}
1 & 1 & 1 \\
1 & 3 & 6 \\
1 & 4 & 2 \\
1 & 6 & 4 \\
1 & 2 & 5
\end{array} \right) \otimes
\left(
\begin{array}{cccccccc}
0 & 0 & 0 & 0 & 0 & 0 & 0 & 1 \\
0 & 0 & 0 & 0 & 0 & 0 & 1 & 0 \\
0 & 0 & 0 & 0 & 0 & 1 & 0 & 0 \\
0 & 0 & 0 & 0 & 1 & 0 & 0 & 0 \\
0 & 0 & 0 & 1 & 0 & 0 & 0 & 0 \\
0 & 0 & 1 & 0 & 0 & 0 & 0 & 0 \\
0 & 1 & 0 & 0 & 0 & 0 & 0 & 0 \\
1 & 0 & 0 & 0 & 0 & 0 & 0 & 0
\end{array}
\right).
\end{align}
Hence it is possible to express the generator matrix in the compact form of \eqref{eq:our-code}.
\end{example}

We finish the code design with the following conclusion regarding the density.
\begin{proposition}
\label{prop:alg-is-LD}
The array code in Algorithm \ref{alg:array_code} is lowest density, \emph{i.e.}, its row and column weight is $r$ and $k$, respectively.
\end{proposition}
\begin{IEEEproof}
Each rowwise MDS code consists of $k$ data symbols and $r$ parity symbols; hence each parity symbol is a function of $k$ data symbols and each data symbol appears in $r$ parity symbols. Consequently the generator matrix $\mbf{A}$ has row and column weight $r$ and $k$, respectively.

This can also be seen from \eqref{eq:our-code}. Since $\tilde{\mbf{A}}$ is a totally nonsingular $k \times r$ matrix, it has no zero entries. Hence a Kronecker product with the ``antidiagonal'' matrix $\mbf{D}$ will produce a generator matrix with column and row weight $k$ and $r$, respectively.
\end{IEEEproof}

\section{Incomplete networks}
\label{sec:incomplete}

In order to study incomplete networks and in which cases lowest density MDS array codes exist for them, we will describe them as graphs \cite{SaT13c}. In this section we study how some of their properties affect the code design.

\subsection{Incomplete graphs}
We begin with some useful definitions and lemmas.
\begin{definition}
A graph (finite, undirected, without loops and multiple edges) is defined as $G=(V,E)$, where $V=V(G)$ is a nonempty set of nodes (vertices) and $E=E(G)$ is a set of edges connecting the nodes.
\end{definition}

\begin{definition}
If all nodes are connected with one another, a graph is said to be \emph{complete}, otherwise it is \emph{incomplete}.
\end{definition}

\begin{definition}
A graph is said to be \emph{regular} if all its nodes have the same degree (number of edges connected to it), otherwise it is said to be \emph{irregular}.
\end{definition}

\begin{definition}
Two graphs $G_1$ and $G_2$ are \emph{isomorphic} if there exists a mapping $\phi: V(G_1) \rightarrow V(G_2)$ such that two nodes $u$ and $v$ in $G_1$ are connected if and only if $\phi(u)$ and $\phi(v)$ are connected in $G_2$.
\end{definition}
Note that two graphs are isomorphic iff the nodes of one graph can be relabelled to obtain the other. It is possible to divide graphs into different classes, where all graphs in a class are isomorphic to one other but two graphs in different classes are nonisomorphic. The number of classes depends on the number of nodes and edges, \emph{e.g.}, $3$-regular graphs with six nodes have two nonisomorphic classes, while $5$-regular graphs with ten nodes have ten nonisomorphic classes \cite{Ste90b}. Note that the concept of isomorphic graphs extends beyond regular graphs; two irregular graphs can be isomorphic, although this requires the distribution of node degrees to be identical. If they are not, no permutation of the nodes can make one graph equal to the other.

\begin{lemma}
\label{lemma:subgraph}
If a lowest density MDS code exists for the graph $G_1$ which is a subgraph of $G_2$ ($E(G_1) \subseteq E(G_2$)) with the same number of nodes, then the code is also valid for $G_2$.
\end{lemma}

\begin{IEEEproof}
An edge in a graph indicates the possibility of the corresponding entries $\mbf{A}_{i,j}$ in the generator matrix to be nonzero; the absence of an edge means that those values must be zero. If $G_1$ is a subgraph of $G_2$ then the code for $G_1$ can be directly applied to $G_2$; the generator matrix entries corresponding to the edges which are in $G_2$ but not $G_1$ can simply be set to zero.
\end{IEEEproof}

This means that if we can find a code for $G_1$, we do not need to look for codes for graphs $G_2$ of which $G_1$ is a subset. Note that we limit our study to connected graphs (every two nodes can be connected through, potentially, other nodes), otherwise we can simply split the graph into subgraphs that are connected.

\subsection{Code design}

In this section we will derive necessary conditions for the existence of lowest density MDS array codes for incomplete graphs. We will also give some explicit examples of codes. We start by looking at the necessary node degree of a graph.

\begin{proposition}
\label{proposition:d-min}
A necessary condition for an MDS code to exist over an incomplete graph is
that the minimum node degree is $d_{min}\geq \max \left\{ k,r\right\} $.
\end{proposition}

\begin{IEEEproof}
Assume $k\geq r$ first. If a node has degree $d$, then there are $n-d$
zero blocks $\mbf{A}_{i,j}$ in a block column. However, only $n-1-d$ such blocks can be part of any of the $n\choose r$ submatrices $\mbf{A}^{(f)}$ because the diagonal blocks $\mbf{A}_{i,i}$ are never included in $\mbf{A}^{(f)}$ by construction. This number must
always be smaller than $r$, otherwise we could select $r$ block rows for ${\cal F}$ yielding
a singular submatrix $\mbf{A}^{(f)}$ with some all-zero columns. This would make it impossible to design an MDS code because of Proposition~\ref{prop:NecAndSuf}. Therefore we obtain
\begin{equation}
n-1-d<r\Rightarrow d>k-1\Rightarrow d\geq k.
\end{equation}%
Now consider the case $k<r$. If a node has degree $d$, then there are $n-1-d$ zero
blocks in a block row that can be part of a submatrix $\mbf{A}^{(f)}$. This must always be
smaller than $k$, otherwise we could select $k$ block rows for ${\cal S}$ and make up a
submatrix who has some all-zero rows and hence is singular. Thus%
\begin{equation}
n-1-d<k\Rightarrow d>r-1\Rightarrow d\geq r.
\end{equation}%
Hence we can conclude that $d_{\min }\geq \max \left\{ k,r\right\} $.
\end{IEEEproof}

The code in \eqref{eq:our-code} has an important structural property that allows it to be applied to certain classes of incomplete graphs.
\begin{lemma}
\label{lemma:our-code-node-degree}
The generator matrix in (\ref{eq:our-code}) has $n - \gcd(k,r)$ nonzero blocks $\mbf{A}_{i,j}$ per block row/column. These are symmetrically arranged such that $\mbf{A}_{i,j}=\mbf{0} \Rightarrow \mbf{A}_{j,i}=\mbf{0}$.
\end{lemma}

\begin{IEEEproof}
See Appendix \ref{app:our-code-node-degree}.
\end{IEEEproof}

We will now study three special cases and discuss the design of lowest density MDS codes for them.

\subsubsection{$r=1$}
When designing codes for $r=1$ failures, we get

\begin{equation}
k = n-1 \Rightarrow m = \frac{k}{\gcd (k,r)} = n-1\;,\; p = \frac{rm}{k} = 1.
\end{equation}
According to Proposition \ref{proposition:d-min}, we then need a graph with minimum node degree $d_{min} = \max \left\{n-1,1\right\} = n-1$. However this is a complete graph, where all nodes are connected to each other. Although this sounds very restrictive, it can be explained by looking at the actual code. There are $n-1$ data symbols and only one parity symbol per node and since for a lowest density MDS code, each data symbol can only appear in $r=1$ parity symbol, the $m=n-1$ data symbols in a node must be spread out to all the other $n-1$ nodes. This requires a complete graph as all nodes must be connected to each other. For complete graphs and $r=1$ failures, we can use a simple parity-bit code \cite{Blaum1996}.

\subsubsection{$r=2$, $n$ even}
In this case, we will show that we can reduce the code design problem to a single graph for which we have an explicit code design. First we start with the problem of reducing graphs.

\begin{lemma}
\label{lemma:graph-r-2}
All graphs with $n$ even nodes and minimum node degree $n-2$ can be
reduced to a regular graph with degree $n-2$ by removing some edges.
\end{lemma}

\begin{IEEEproof}
Since the minimum node degree is $n-2$, each node must have a degree of
either $n-2$ or $n-1$. Assume that there are $l$ nodes with degree $n-1$ and
$n-l$ nodes with degree $n-2$. The number of edges is then%
\begin{equation}
\frac{l\left( n-1\right) +\left( n-l\right) \left( n-2\right) }{2}=\frac{%
n^{2}-2n+l}{2}
\end{equation}%
and since $n$ is even, so is $l$. These $l$ nodes with degree $n-1$ form a \emph{clique}, \emph{i.e.}, a complete subgraph where every pair of nodes are connected by an edge, hence we can remove $l/2$ edges between as many nonoverlapping pairs without affecting the minimum node degree. In the resulting graph all nodes have the same degree $n-2$, which proves the assert.
\end{IEEEproof}

We can now turn our attention to the $(n-2)$-regular graphs and show that they form a single nonisomorphic class. Firstly, we recall a couple of definitions in graph theory. A \emph{complement graph}, $G_2$, of a graph $G_1$ is a graph on the same nodes such that two nodes in $G_2$ are adjacent if they are not adjacent in $G_1$. In other words, the \emph{sum} of a graph with its complement produces a complete graph. A \emph{perfect matching} of a graph is a 1-regular graph on the same nodes, \emph{i.e.}, a graph whose edges match all nodes but have no node in common.

\begin{lemma}
\label{lemma:one-isomorphic-class}
All $\left( n-2\right) $-regular graphs with $n$ even nodes are
isomorphic, \emph{i.e.}, there is only one nonisomorphic class.
\end{lemma}

\begin{IEEEproof}
Since all nodes have degree $n-2$, they are connected to all other nodes
except one. Hence, the complement graph forms a perfect matching because all nodes have
exactly one missing edge. Because all the $(n-2)$-regular graphs with $n$ nodes can be constructed by removing a perfect matching from a complete graph with $n$ nodes, we can find the total number of $(n-2)$-regular graphs by counting the number of distinct perfect matchings of $n$ nodes. Each perfect matching is obtained by dividing the $n$ nodes into two
sets of size $n/2$ and matching each node in one set with a node in the other
set. Dividing the nodes into two sets can be done in $\binom{n}{n/2}$ ways
and then each set can be reordered in $\left( n/2\right) !$ ways; hence the
total number of perfect matchings is%
\begin{equation}
\binom{n}{n/2}\times \left( \frac{n}{2}\right) !\times \left( \frac{n}{2}%
\right) !=\frac{n!\left( n/2\right) !\left( n/2\right) !}{\left( n/2\right)
!\left( n/2\right) !}=n!.
\end{equation}%
Since the nodes can be permuted in $n!$ ways, this leaves only one possible
graph when permutations are discounted. Hence there is only one
nonisomorphic class.
\end{IEEEproof}

This leads us to stating the main result in this subsection.
\begin{proposition}
The code in (\ref{eq:our-code}) is a $[n,k=n-2]$ lowest density MDS array code for all graphs with even $n$ and minimum node degree $n-2$.
\end{proposition}

\begin{IEEEproof}
The data and parity symbols per node are
\begin{align}
m &= \frac{k}{\gcd(k,r)}=\frac{n}{2}-1\\
p &= \frac{r}{\gcd(k,r)}=1. \label{eq:p_r2}
\end{align}
From Propositions~\ref{prop:alg-is-MDS} and \ref{prop:alg-is-LD} we know that the code (\ref{eq:our-code}) is lowest density MDS. According to Lemma~\ref{lemma:our-code-node-degree} the number of nonzero blocks $\mbf{A}_{i,j}$ per row and column is $n-\gcd(k,r) = n-2$ since $k$ is an even number. This means that for each row $i$ there is a block in column $j$ (apart from the diagonal) which satisfies $\mbf{A}_{i,j} = \mbf{A}_{j,i} = \mbf{0}$. This implies that there does not have to be any connection between nodes $i$ and $j$ and each node only needs to be connected to $n-2$ other nodes. Therefore, the code is guaranteed to work on any $(n-2)$-regular graph with $n$ (even) nodes because there is only one nonisomorphic class according to Lemma \ref{lemma:one-isomorphic-class}. Finally, from Lemma \ref{lemma:graph-r-2} we know that all graphs with even $n$ and minimum node degree $n-2$ can be reduced to a $(n-2)$-regular graph, which together with Lemma~\ref{lemma:subgraph} proves the assert.
\end{IEEEproof}

As a remark, we note that the so-called B-code in \cite{XBB99a} is also capable of correcting $r=2$ erasures out of $n$ (even) nodes. However the generator matrix of the B-code would require that all nodes are connected to each other, \emph{i.e.}, it would not work for an incomplete graph. The same is true for the codes in \cite{CaB09a,JJF09c}.

\begin{figure}%
\centering
\includegraphics[width=0.9\linewidth]{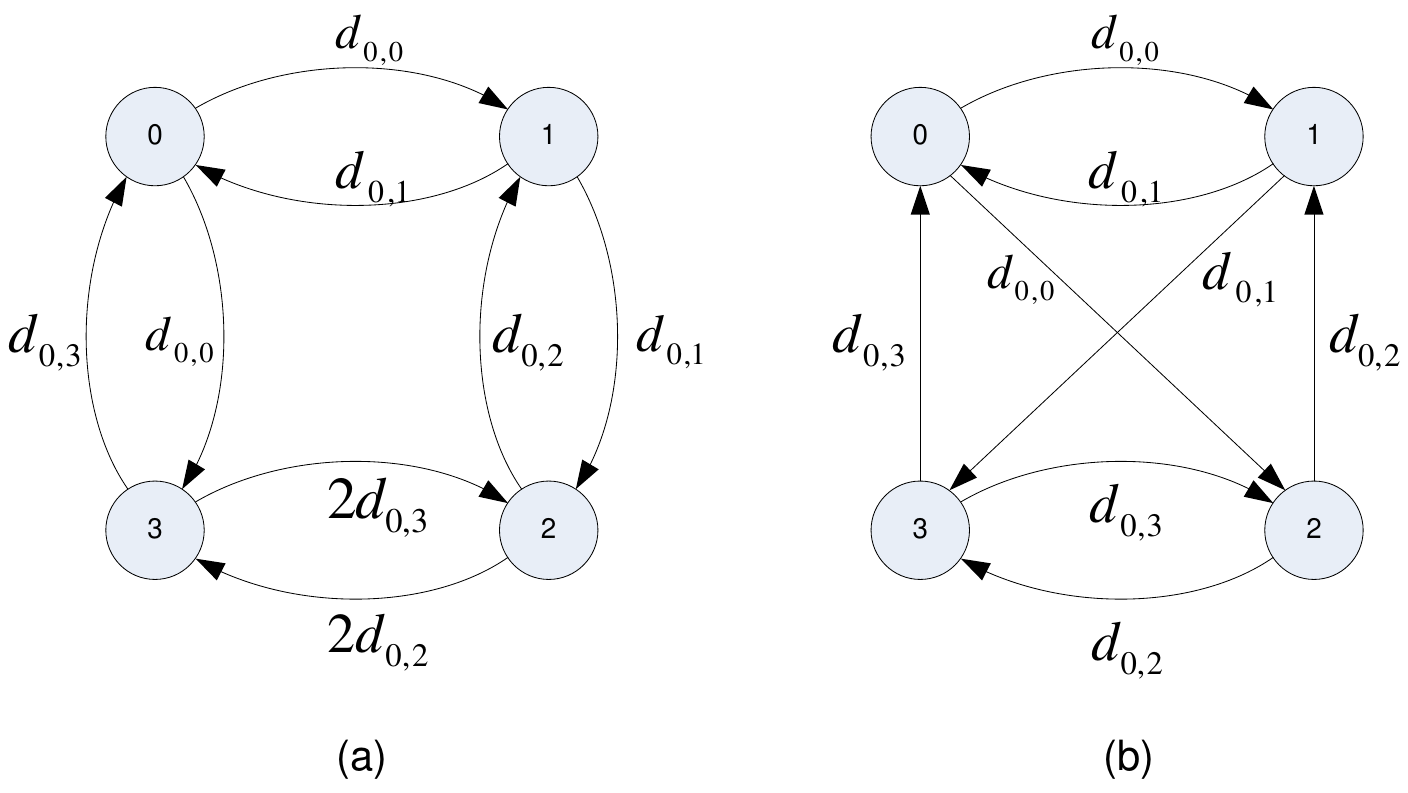}
\caption{Two lowest density MDS array codes for $n=4$ and $r=2$: (a) the code (\ref{eq:our-code}) works on an incomplete graph, (b) the code \cite{XBB99a} requires a complete graph.}%
\label{fig:code-comparison}%
\end{figure}

\begin{example}
Consider $n=4$ and $r=2$. Hence, $m=p=1$. By using Algorithm~\ref{alg:array_code}, we can construct a $[4,2]$ lowest density MDS array code by independently encoding each of the two rows of the following array
\begin{equation}
\vphantom{
    \begin{array}{c}
    \overbrace{XYZ}^{\mbox{$R$}}\\ \\
    \underbrace{pqr}_{\mbox{$S$}}
    \end{array}}%
\begin{array}{c}
\coolleftbraceone{[4,2]}{0}\\
\coolleftbraceone{[4,2]}{0}
\end{array}%
\left(\begin{array}{cccc}
\cooloverone{0}{d_{0,0}} & \cooloverone{1}{f_{0,1}} & \cooloverone{2}{d_{0,2}} & \cooloverone{3}{f_{0,3}}\\ \hline
f_{0,0} & d_{0,1} & f_{0,2} & d_{0,3}
\end{array}\right)%
\end{equation}
By using a Singleton matrix \cite{RoS85a} as the totally nonsingular matrix, the generator matrix for the code in \eqref{eq:our-code} is
\begin{equation}
\label{eq:our-code-n4k2}
\mbf{A} = \left(
\begin{array}{cc}
1 & 1 \\
1 & 2
\end{array}
\right) \otimes
\left(
\begin{array}{cc}
0 & 1 \\
1 & 0
\end{array}
\right) = \left(
\begin{array}{cccc}
0 & 1 & 0 & 1 \\
1 & 0 & 1 & 0 \\
0 & 1 & 0 & 2 \\
1 & 0 & 2 & 0
\end{array}
\right)
\end{equation}
whereas the generator matrix for the B-code is \cite{XBB99a}
\begin{equation}
\label{eq:B-code-n4k2}
\mbf{A}_{\text{B-code}} = \left(
\begin{array}{cccc}
0 & 1 & 1 & 0 \\
1 & 0 & 0 & 1 \\
0 & 1 & 0 & 1 \\
1 & 0 & 1 & 0
\end{array}
\right).
\end{equation}
Since $m=p=1$, it is clear from (\ref{eq:B-code-n4k2}) that all nodes need to be connected to each other to exchange information. On the other hand, from (\ref{eq:our-code-n4k2}) we see that nodes 0 and 2 do not need to be connected (nor nodes 1 and 3); hence this code will work for incomplete graphs with minimum node degree two. This is illustrated in Figure \ref{fig:code-comparison}. Lemma \ref{lemma:one-isomorphic-class} ensures that there is only one nonisomorphic class and hence the code will work for all graphs of this type.
\end{example}

\subsubsection{$n$ divisible by $r$}

If $n$ is divisible by $r$, we can use the results obtained above to prove the following assert.
\begin{proposition}
For any pair of parameter values $[n,r]$ with $r\,|\,n$, the code in (\ref{eq:our-code}) is a lowest density MDS array code for one nonisomorphic class of a regular graph with $n$ nodes and node degree $d=\max\{r,n-r\}$, as well as all graphs of which it is a subgraph.
\end{proposition}
\begin{IEEEproof}
If $n$ is divisible by $r$, then $k=n-r$ is also divisible by $r$. Hence, from Lemma~\ref{lemma:our-code-node-degree}, the generator matrix in \eqref{eq:our-code} has only $n-\gcd\{k,r\} = n-\min \{k,r\} = \max \{k,r\}$ nonzero blocks per row/column. This means that the code, which is lowest density MDS by Propositions~\ref{prop:alg-is-MDS} and \ref{prop:alg-is-LD}, could be used for a regular graph with node degree $d = \max\{k,r\}$. However, there are in general many nonisomorphic classes of regular graphs for $r > 2$, so the code can only be applied to one of these classes. For example, there are four nonisomorphic classes of a 6-regular graph with $n=9$ nodes. Moreover, because of Lemma~\ref{lemma:subgraph}, the code is also valid for graphs of which any representative of the regular nonisomorphic class is a subgraph.
\end{IEEEproof}

\subsubsection{General graphs}
\begin{figure}[htb]%
\centering
\subfloat[]{\includegraphics[width=0.45\linewidth]{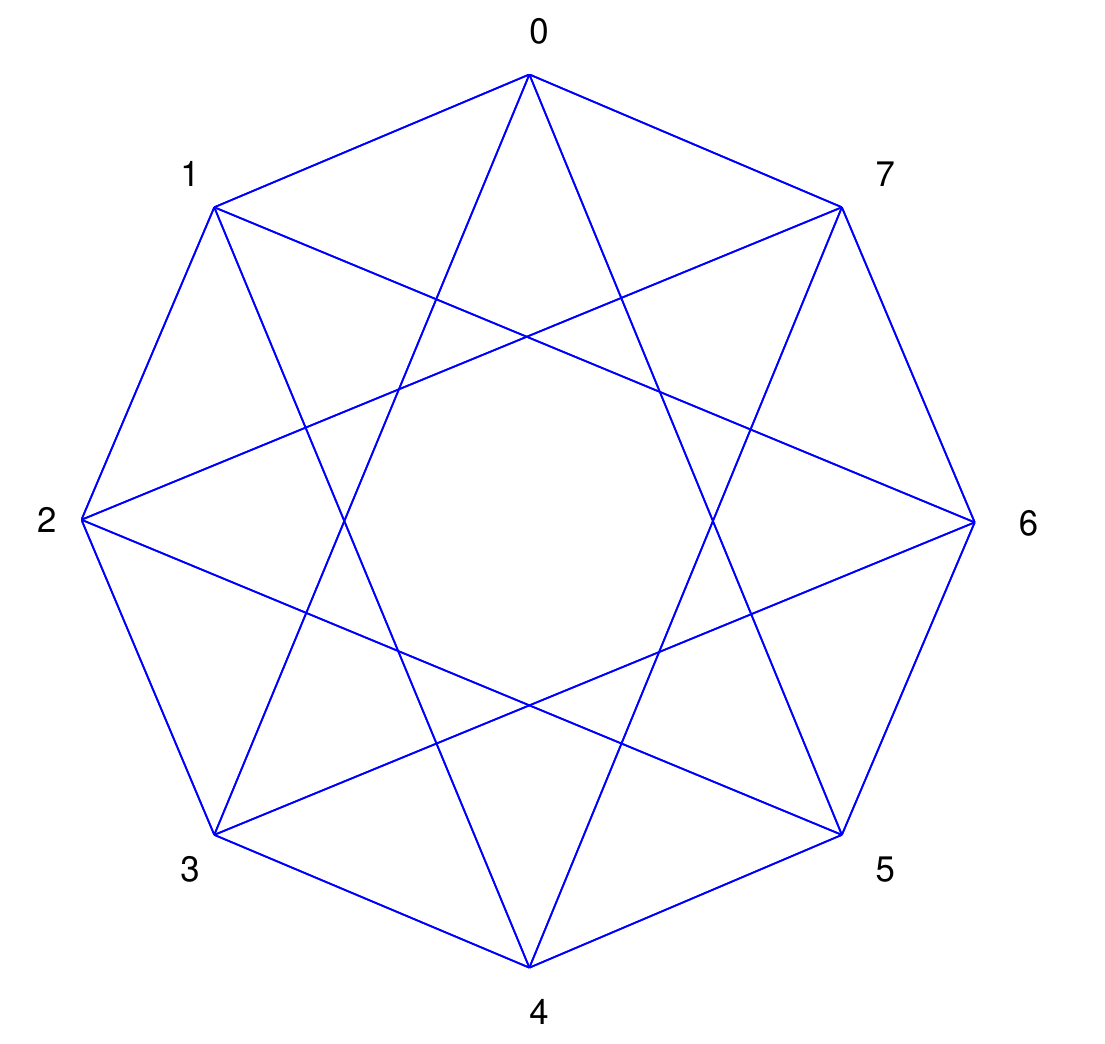}}%
\subfloat[]{\includegraphics[width=0.45\linewidth]{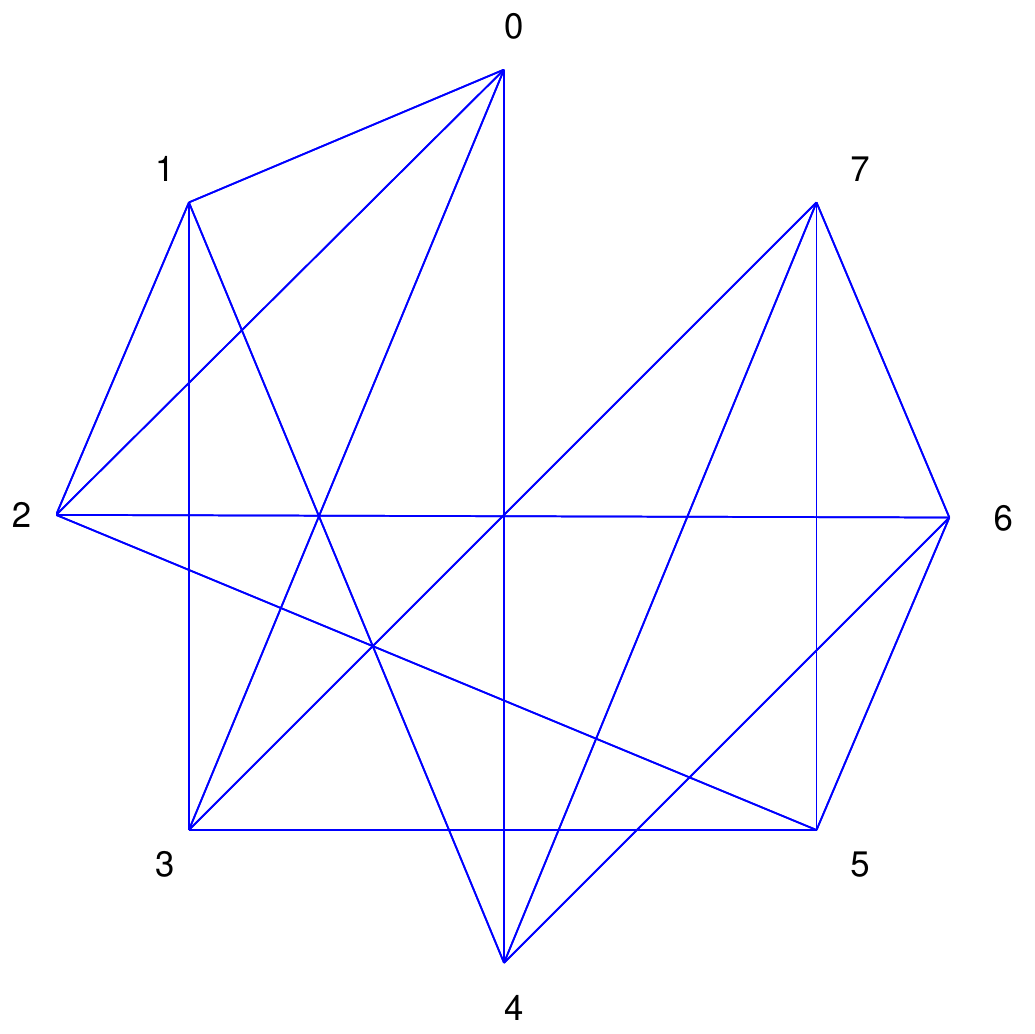}}
\caption{The 4-regular graph in (a) allows for a lowest density MDS code whereas the graph in (b) does not.}%
\label{fig:reg_n8k4}%
\end{figure}

For general graphs satisfying Proposition~\ref{proposition:d-min}, other than the special cases discussed in the previous subsections, it is very difficult to say whether or not a lowest density MDS code exists. For instance, assume that we seek to design an $[n=8, k=4]$ code (and hence $p=m=1$) on a 4-regular graph. There are six nonisomorphic 4-regular graphs with $n=8$ nodes and from the previous subsection, we know that for one of them we can use \eqref{eq:our-code} to design an $[n=8,k=4]$ lowest density MDS code; this graph is shown in Figure \ref{fig:reg_n8k4}a. However for the graph in Figure \ref{fig:reg_n8k4}b and all the graphs isomorphic to it, no lowest density MDS code exists. This can be shown by considering the failure pattern ${\cal F} = \left\{0,1,2,3\right\}$ (and consequently ${\cal S} = \left\{4,5,6,7\right\}$), which means that the submatrix is

\begin{equation}
\mbf{A}^{(f)} = \left(
\begin{array}{cccc}
A_{0,4} & 0 & 0 & 0 \\
A_{1,4} & 0 & 0 & 0 \\
0 & A_{2,5} & A_{2.6} & 0 \\
0 & A_{3,5} & 0 & A_{3,7}
\end{array}
\right).
\end{equation}
It is clear that the first two rows are linearly dependent regardless of what $A_{0,4}$ and $A_{1,4}$ are. Hence $\mbf{A}^{(f)}$ is always singular, which means that it is impossible to design a MDS code for this graph (see Proposition~\ref{prop:NecAndSuf}). This example illustrates the difficulties in determining if a lowest density MDS array code exists for a given incomplete network.

\section{Conclusion}
\label{sec:conclusions}

In this paper we have introduced a lowest density MDS array code and applied it to Smart Meter networks. By adding redundancy, \emph{i.e.}, each node in the network also stores some parity symbols relating to other nodes, the network can be made more reliable and recover from outages. A concentrator node can collect the whole network data even if only a subset of the nodes are available. This is achieved by using a vertical MDS array code with lowest density, which is capable of correcting $r$ erasures out of $n$ nodes. Unlike existing lowest density MDS array codes, which either have prime number related restrictions on $n$ and/or only works for specific (low) number of failures $r$, the proposed encoding has no restrictions on $n$ or $r$ so it can be constructed for any number of nodes and any number of failures. It is a simple design involving finding the optimal size of the array, arranging the data and parity symbols appropriately in the array and then applying stacked MDS codes. We also studied incomplete networks where all nodes are not connected to each other. By using graph theory, we derived conditions on the minimum node degree for the existence of lowest density MDS array codes and also provided an explicit design for two node failures in networks with an even number of nodes.

\section*{Acknowledgement}

The authors would like to acknowledge the fruitful discussions with their colleagues at Toshiba Research Europe Ltd. and the support of its directors.

\appendices

\section{Proof of Proposition \ref{prop:stacked_code}}
\label{app:stacked_code}

The order of the Kronecker product (\ref{eq:our-code}) can be reversed by means of the \emph{commutation matrix}. By using the property \cite[Th.~3.1-(i) and (viii)]{Magnus1979} we can write
\begin{equation}
\mbf{A} = \tilde{\mbf{A}} \otimes \mbf{D} = \mbf{K}_{k(m+p)} (\mbf{D} \otimes \tilde{\mbf{A}}) \mbf{K}_{(m+p)r}
\label{eq:kron_comm}
\end{equation}
where the commutation matrices $\mbf{K}_{k(m+p)}$ and $\mbf{K}_{(m+p)r}$ can be represented as
\begin{align}
\mbf{K}_{k(m+p)} &= \sum_{i=0}^{m+p-1}{\mbf{e}_{i} \otimes \mbf{I}_{k} \otimes \mbf{e}_{i}^T}\\
\mbf{K}_{(m+p)r} &= \sum_{i=0}^{m+p-1}{\mbf{e}_{i} \otimes \mbf{I}_{r} \otimes \mbf{e}_{i}^T}
\end{align}
where $\mbf{e}_i$ is the $i$th row unit vector of order $(m+p)$.

Let us introduce the following notation. Since vectors in this paper are row vectors, for consistency we indicate with $\vecbar{\cdot}$ the transposed vector operator, \emph{i.e.}, the linear transformation of a matrix into a row vector. We also denote by $(\mbf{X})_{ab}$ the reshaping operation on matrix $\mbf{X}$, obtained by reading the elements of $\mbf{X}$ column-wise and rearranging them in an $a\times b$ array. Let us partition the array (\ref{eq:array}) in a data and parity submatrix as follows
\begin{equation}
\left(\begin{array}{c}
\mbf{C}_d \\ \hline
\mbf{C}_f
\end{array}\right)%
\end{equation}
where $\mbf{C}_d$ is $m\times n$ and $\mbf{C}_f$ is $p\times n$. Because the code parameters fulfill (\ref{prop:pk_mr}), we obtain
\begin{align}
mn &= \frac{kp}{r}(k+r) = k(m+p) \\
pn &= \frac{rm}{k}(k+r) = r(m+p)
\end{align}
hence we can write the data and parity vectors of (\ref{eq:c_notation1})-(\ref{eq:dG}) as
\begin{equation}
\mbf{d} =\vecbar{(\mbf{C}_d)_{(m+p)k}}\quad\textrm{and}\quad \mbf{f}=\vecbar{(\mbf{C}_f)_{(m+p)r}}. \label{eq:d_f}
\end{equation}

By using (\ref{eq:kron_comm}), we can rewrite the nonsystematic part of (\ref{eq:dG}) as
\begin{equation}
\mbf{f}= \mbf{d} \mbf{K}_{k(m+p)} (\mbf{D} \otimes \tilde{\mbf{A}}) \mbf{K}_{(m+p)r}
\end{equation}
and because $\mbf{K}_{(m+p)r}\mbf{K}_{r(m+p)}=\mbf{I}_{r(m+p)}$ (see \cite[Th.~3.1-(ii) and (iii)]{Magnus1979}), we obtain
\begin{equation}
\mbf{f}\mbf{K}_{r(m+p)} = \mbf{d} \mbf{K}_{k(m+p)} (\mbf{D} \otimes \tilde{\mbf{A}}).
\label{eq:f_d_eq}
\end{equation}
As the commutation matrix $\mbf{K}_{ba}$ is such that $\vecbar{\mbf{X}}\mbf{K}_{ba} = \vecbar{\mbf{X}^T}$, where $\mbf{X}$ is an $a\times b$ matrix (this property is equivalent to \cite[Th.~3.1-(vii)]{Magnus1979}) and after replacing (\ref{eq:d_f}) in (\ref{eq:f_d_eq}), we get
\begin{equation}
\vecbar{(\mbf{C}_f)_{(m+p)r}^T} = \vecbar{(\mbf{C}_d)_{(m+p)k}^T} (\mbf{D} \otimes \tilde{\mbf{A}}).
\end{equation}
By using basic properties of the vector operator, and particularly the fact that $\vecbar{\mbf{ABC}} = \vecbar{\mbf{B}}(\mbf{C}\otimes \mbf{A}^T)$, we derive
\begin{multline}
\vecbar{(\mbf{C}_f)_{(m+p)r}^T\mbf{D}} = \vecbar{(\mbf{C}_f)_{(m+p)r}^T} (\mbf{D}\otimes\mbf{I}_r) \\
= \vecbar{(\mbf{C}_d)_{(m+p)k}^T} (\mbf{D} \otimes \tilde{\mbf{A}}) (\mbf{D}\otimes\mbf{I}_r)\\
= \vecbar{(\mbf{C}_d)_{(m+p)k}^T} (\mbf{I}_{m+p} \otimes \tilde{\mbf{A}}).
\label{eq:almost_final}
\end{multline}
After introducing the vectors
\begin{align}
\mbf{d}' &=[\mbf{d}'_0,\dots,\mbf{d}'_{m+p-1}]\\
\mbf{f}' &=[\mbf{f}'_0,\dots,\mbf{f}'_{m+p-1}]
\end{align}
as shuffled versions of $\mbf{d}$ and $\mbf{f}$, respectively, we can rewrite (\ref{eq:almost_final}) as a set of $m+p$ equations
\begin{equation}
\mbf{f}'_i = \mbf{d}'_i \tilde{\mbf{A}},\quad i=0,\dots,m+p-1
\end{equation}
which describe the encoding of an $[n,k,d=n-k+1]$ generalised RS code in $\F_q$ \cite{RoS85a}. Therefore, the code (\ref{eq:our-code}) is equivalent to independently encoding each row of the array (\ref{eq:array}), after an appropriate rearrangement of the elements on each column, with a linear MDS code. The array code is MDS because obtained by stacking $m+p$ MDS codewords. It also meets the density bounds of Proposition \ref{prop:lowest_density} with equality because every parity symbol depends only on the $k$ data symbols in its row and every data symbol only contributes to the $r$ parity symbols of its row, hence the code is lowest density.

Finally, we need to show that all the data and parity symbols in the $j$th column can be mapped consistently in the $j$th column of the shuffled array. From (\ref{eq:almost_final}) it is not difficult to see that, after denoting the linearised index of data symbols $d_{i,j}$ as $l=i+mj$, its new row location in the rearranged array is given by $l\mod(m+p)$, for $i=0,\dots,m-1$. Similarly, for parity symbol $f_{\iota,j}$, with linearised index $t=\iota+pj$, after including the effect of matrix $\mbf{D}$ acting on the columns of $(\mbf{C}_f)_{(m+p)r}^T$, the new row location is given by $(m+p-1-t)\mod(m+p)=(mj-\iota-1)\mod(m+p)$ for $\iota=0,\dots,p-1$.

We can easily verify that the two new row locations never overlap by checking that the following equation has no solution in the range of indices $i$ and $\iota$
\begin{align}
& (i+mj)\!\!\!\!\mod(m+p) = (mj-\iota-1)\!\!\!\!\mod(m+p) \notag\\
\Rightarrow & (i+\iota+1)\mod(m+p) = 0.
\end{align}
The latter has no solution because $1\leq i+\iota+1\leq m+p-1$ for $i=0,\dots,m-1$ and $\iota=0,\dots,p-1$.

\section{Proof of Lemma \ref{lemma:our-code-node-degree}}
\label{app:our-code-node-degree}

We can explicitly find which blocks of (\ref{eq:our-code}) are zero. First note that $i+j+1\bmod (m+p)=0$ if the element $A_{i,j} $ is
nonzero, which means that we can count the number of nonzero elements in an arbitrary block
in position $\left( i^{\prime },j^{\prime }\right) $ as
\begin{align}
& \sum_{i=0}^{m-1}\sum_{j=0}^{p-1}\delta \left[ i^{\prime }m+i+j^{\prime
}p+j+1\bmod (m+p)\right] \nonumber \\
= & \!\!\sum_{i=0}^{m-1}\!\sum_{j=0}^{p-1}\!\delta \left[ i^{\prime }\!\!\left(
m+p\right)\! +\!\left( j^{\prime }-i^{\prime }\right) p +i+j+1\bmod\! (m+p) \right] \nonumber \\
= & \sum_{i=0}^{m-1}\sum_{j=0}^{p-1}\delta \left[ \left( j^{\prime
}-i^{\prime }\right) p+i+j+1\bmod (m+p) \right] .
\end{align}%
If this sum is nonzero, we must have at least one pair $\left( i,j\right) $
such that%
\begin{equation}
\left( j^{\prime }-i^{\prime }\right) p+i+j+1\bmod (m+p) =0.\label{eq:sum-block1}
\end{equation}
By using the property
\begin{equation}
(a+b)\bmod c = 0\;\Rightarrow\; b\bmod c = (-a)\bmod c
\label{eq:mod_equa}
\end{equation}
we obtain
\begin{equation}
i+j+1\bmod (m+p)=\left( i^{\prime }-j^{\prime }\right) p \bmod (m+p) \label{eq:sum-block2}\\
\end{equation}
and, after noting that%
\begin{equation}
1\leq i+j+1\leq m+p-1.
\label{eq:inequa}
\end{equation}
we can write
\begin{equation}
i+j+1=\left( i^{\prime }-j^{\prime }\right) p \bmod (m+p). \label{eq:sum-block3}
\end{equation}%

Therefore, if $\left( i^{\prime }-j^{\prime }\right) p \bmod (m+p)=0$, it
follows from (\ref{eq:inequa}) that (\ref{eq:sum-block3}) has no solution, hence the block $\left(
i^{\prime },j^{\prime }\right) $ contains all zeros. On the other hand, if $%
\left( i^{\prime }-j^{\prime }\right) p\bmod (m+p) \neq 0$, there exist index pairs $%
\left( i,j\right) $ which fulfill (\ref{eq:sum-block3}), since $i+j+1$ can
attain any value between $1$ and $m+p-1$; in this case the block $\left(
i^{\prime },j^{\prime }\right) $ is nonzero. Trivially, for $i^{\prime
}=j^{\prime }$ we have $\left( i^{\prime }-j^{\prime }\right) p \bmod
(m+p)=0$ such that the diagonal blocks are all zeros. Since $m$ and $p$ are
coprime, $p$ and $m+p$ are also coprime; this means that $i^{\prime
}-j^{\prime }$ must be a multiple of $m+p$ in order to have: $\left( i^{\prime
}-j^{\prime }\right) p \bmod (m+p)=0$. From \eqref{eq:MDS_necessary} we obtain
\begin{equation}
m+p = \frac{n}{\gcd(k,r)}
\end{equation}
and thus
\begin{equation}
\left\vert i^{\prime }-j^{\prime }\right\vert \leq n-1=\gcd \left(
k,r\right) \left( m+p\right) -1.
\end{equation}
We see that there are $\gcd \left( k,r\right) -1$ values of $\left\vert
i^{\prime }-j^{\prime }\right\vert $ that are multiples of $m+p$ (excluding
the diagonal block with $i^{\prime }-j^{\prime }=0$). This means that in each block row
there will be $\gcd \left( k,r\right) -1$ off-diagonal zero blocks.
Hence the number of nonzero blocks is $n-1-\left( \gcd \left( k,r\right)
-1\right) =n-\gcd \left( k,r\right) $.

The above analysis only shows that each row and column has $n-\gcd(k,r)$ nonzero blocks. We also need to show that the generator matrix $\mbf{A}$ is block symmetric in the sense that $\mbf{A}_{i',j'} = \mbf{0} \Rightarrow \mbf{A}_{j',i'} = \mbf{0}$. This property is required for $\mbf{A}$ to be associated to a graph with minimum degree $n-\gcd(k,r)$. From

\begin{align}
\mbf{A}_{i',j'} = \mbf{0} \Leftrightarrow & \hspace*{2mm} A_{i'm+i,j'p+j} = 0 \\
\Leftrightarrow & \hspace*{2mm}  i'm+i+j'p+j+1 \bmod (m+p) \neq 0 \label{eq:allzeros1}\\
& \forall i \in \left[0,m-1\right],~ \forall j \in \left[ 0,p-1 \right] \nonumber
\end{align}
by subtracting $(i'+j'+1)(m+p)$ from the left-hand side of (\ref{eq:allzeros1}), we obtain
\begin{equation}
-i'p+i-j'm+j+1 - (m+p) \bmod (m+p) \neq 0.
\label{eq:allzeros2}
\end{equation}
By using the property (\ref{eq:mod_equa}) with $b=0$ we can rewrite (\ref{eq:allzeros2}) as
\begin{equation}
i'p+(m-i)+j'm+(p-j)-1 \bmod (m+p) \neq 0
\end{equation}
from which we finally get
\begin{multline}
i'p+(m-1-i)+j'm+(p-1-j)+1 \bmod (m+p) \neq 0\\
\shoveleft{\Rightarrow\quad A_{j'm+(m-1-i),i'p+(p-1-j)} = 0} \\
\forall i \in \left[0,m-1\right],~ \forall j \in \left[0,p-1\right].
\end{multline}
Consequently, if all elements in $\mbf{A}_{i',j'}$ are zero, so are the elements in $\mbf{A}_{j',i'}$ which proves the assertion.

\clearpage

\begin{algorithm}
\caption{Algorithm for designing a $[n,k]$ MDS array code.}
\label{alg:array_code}
\begin{enumerate}
\item Given the number of nodes $n$ and the maximum number of correctable erasures $r=n-k$, assign $m=\frac{k}{\gcd(k,r)}$ data symbols and $p=rm/k$ parity symbols to each node.
\item Consider an $(m+p) \times n$ array where data symbol $d_{i,j}$ is located in row $(i+jm) \mod (m+p)$, column $j$ and parity symbol $f_{i,j}$ is located in row $(jm-i-1) \mod (m+p)$, column $j$.
\item Each row of the array is then encoded with a $[n,k]$ systematic MDS code; the $n-k$ parity symbols can be formed by multiplying the $k$ data symbols with the $(n-k) \times k$ MDS generator matrix, \emph{e.g.}, a Singleton matrix \cite{RoS85a}.
\end{enumerate}
\end{algorithm}

\clearpage

\clearpage

\bibliographystyle{IEEEtran}
\bibliography{IEEEabrv,MDS_array_code}

\end{document}